\documentclass[sigconf]{acmart}
\usepackage{multirow}
\usepackage{color}
\usepackage{colortbl}
\usepackage{stfloats}
\usepackage{longfbox} % for drawing frames on stakeholder aliases
\usepackage{subcaption}
\usepackage{enumitem}
\usepackage{graphicx}
\usepackage{subcaption}
\usepackage{tabularx}
\usepackage{longtable}

\makeatletter
\newdimen\@tempdimd
\makeatother

\definecolor{quotebackground}{HTML}{EFEFEF}
\definecolor{tableheader}{HTML}{EFEFEF}
\definecolor{tablegrayline}{HTML}{e0e0e0}

\newfboxstyle{scenarioparam}{padding=8pt, margin-top=3pt, margin-bottom=3pt, border-style=none, background-color=quotebackground}

\definecolor{minimalcolor}{HTML}{D3E593}
\definecolor{mildcolor}{HTML}{F7C38B}
\definecolor{severecolor}{HTML}{F2A19C}
\newfboxstyle{patternparam}{padding=2pt, border-style=none, height=6pt, border-radius=2pt}

\usepackage{newfloat}% added <<<<
\DeclareFloatingEnvironment[fileext=fml,placement={H},name=Dialogue]{dialogue}
\newenvironment{quotetable}{
\vspace{3mm}
\noindent
\small\sffamily\centering
\setlength{\tabcolsep}{0.25em}
\tabularx{\columnwidth}{@{}p{0.12\columnwidth}>{\raggedright\let\newline\\\arraybackslash\hspace{0pt}}X@{}}
\hline}
{
\arrayrulecolor{black}\hline
\endtabularx
\vspace{3mm}
}

\newcommand{\quotebotline}[1]{\textbf{PITCH} & \textit{#1}\\\arrayrulecolor{tablegrayline}\hline}

\newcommand{\quoteuserline}[2]{\textbf{#1} & 
\textit{#2} \\\arrayrulecolor{tablegrayline}\hline}

\newcommand{\quoteomit}[1]{%
  \textbf{} & \textit{#1}\\\arrayrulecolor{tablegrayline}
}

\AtBeginDocument{%
  }

\copyrightyear{2026}
\acmYear{2026}
\setcopyright{cc}
\setcctype{by}
\acmConference[CHI '26]{Proceedings of the 2026 CHI Conference on Human Factors in Computing Systems}{April 13--17, 2026}{Barcelona, Spain}
\acmBooktitle{Proceedings of the 2026 CHI Conference on Human Factors in Computing Systems (CHI '26), April 13--17, 2026, Barcelona, Spain}
\acmPrice{}
\acmDOI{10.1145/3772318.3790957}
\acmISBN{979-8-4007-2278-3/2026/04}

\newcommand{\sys}{PITCH}

\newcommand{\out}[1]{} % Use this line and uncomment it if you want to hide all the in-line comments.

\newcommand{\revision}[1]{{\color{black} #1}}
\newcommand{\revisionCHI}[1]{{\color{black} #1}}

\let\oldquote\quote
\renewcommand\quote{\setlength{\textwidth}{2cm}\oldquote}%%

\begin{document}

\title[Understanding How Users Engage with a Proactive Agent for Daily Planning and Self-Reflection]{\emph{``Having Lunch Now''}: Understanding How Users Engage with a Proactive Agent for Daily Planning and Self-Reflection}

\author{Adnan Abbas}
\orcid{0009-0005-8728-875X}
\affiliation{%
  \institution{Virginia Tech}
  \city{Blacksburg}
  \state{VA}
  \country{USA}
}
\email{adnana99@vt.edu}

\author{Caleb Wohn}
\orcid{0009-0005-2475-5891}
\affiliation{
  \institution{Virginia Tech}
  \city{Blacksburg}
  \state{VA}
  \country{USA}
}
\email{cwohn@vt.edu}

\author{Arnav Jagtap}
 \orcid{0009-0000-3238-5492}
\affiliation{
  \institution{Virginia Tech}
  \city{Blacksburg}
  \state{VA}
  \country{USA}
}
\email{arnavj03@vt.edu}

\author{Eugenia H. Rho}
\orcid{0000-0002-0961-4397}
\affiliation{
  \institution{Virginia Tech}
  \city{Blacksburg}
  \state{VA}
  \country{USA}
}
\email{eugenia@vt.edu}

\author{Young-Ho Kim}
\orcid{0000-0002-2681-2774}
\affiliation{%
  \institution{NAVER AI Lab}
  \city{Seongnam}
  \country{Republic of Korea}
}
\email{yghokim@younghokim.net}

\author{Sang Won Lee}
\orcid{0000-0002-1026-315X}
\authornote{Sang Won Lee conducted this work as a visiting scholar at NAVER AI Lab.}
\affiliation{%
  \institution{Virginia Tech}
  \city{Blacksburg}
  \state{VA}
  \country{USA}
}
\email{sangwonlee@vt.edu}

\renewcommand{\shortauthors}{Abbas et al.}

\begin{abstract}

Conversational agents have been studied as tools to scaffold planning and self-reflection for productivity and well-being. 
While prior work has demonstrated positive outcomes, we still lack a clear understanding of what drives these results and how users behave and communicate with agents that act as coaches rather than assistants. 
Such understanding is critical for designing interactions in which agents foster meaningful behavioral change. 
We conducted a 14-day longitudinal study with 12 participants using a proactive agent that initiated regular check-ins to support daily planning and reflection. 
Our findings reveal diverse interaction patterns: participants accepted or negotiated suggestions, developed shared mental models, reported progress, and at times resisted or disengaged. 
We also identified problematic aspects of the agent's behavior, including rigidity, premature turn-taking, and overpromising. 
Our work contributes to understanding how people interact with a proactive, coach-like agent and offers design considerations for facilitating effective behavioral change. 

\end{abstract}


%     \begin{CCSXML}
%    <concept>
%        <concept_id>10010147.10010178.10010179.10010181</concept_id>
%        <concept_desc>Computing methodologies~Discourse, dialogue and pragmatics</concept_desc>
%        <concept_significance>300</concept_significance>
%        </concept>
%    <concept>
%        <concept_id>10003120.10003121.10003124.10010870</concept_id>
%        <concept_desc>Human-centered computing~Natural language interfaces</concept_desc>
%        <concept_significance>300</concept_significance>
%        </concept>
%    <concept>
%        <concept_id>10003120.10003121.10011748</concept_id>
%        <concept_desc>Human-centered computing~Empirical studies in HCI</concept_desc>
%        <concept_significance>300</concept_significance>
%        </concept>
%  </ccs2012>

% \ccsdesc[300]{Computing methodologies~Discourse, dialogue and pragmatics}
% \ccsdesc[300]{Human-centered computing~Natural language interfaces}
% \ccsdesc[300]{Human-centered computing~Empirical studies in HCI}

% Conversational Agents, Productivity, Well-being, LLM, 

%\received{23 February 2025}

\maketitle

\section{Introduction}\label{sec:introduction}

With the presence of large-language models (LLMs), Conversational Agents (CAs) are becoming more human-like, exhibit better conversational intelligence and novelty in interactions~\cite{CueingTheCrowd, ConversationalAbility, sorin2024large}. 
LLM-powered CAs can interpret intent, adapt their tone, and maintain contextual continuity across multiple sessions~\cite{mahmood2025user, ComPeer}. 
Leveraging these capabilities, researchers have studied coaching-style agents: systems designed to advise, persuade, and guide everyday behavior~\cite{GPTCoach, ComPeer, CoachingCoPilot, LanceRhoCommCoach}. 
Such agents are often \emph{proactive}: rather than waiting for user input, they initiate interactions by reminding, suggesting, or steering users toward specific actions~\cite{ComPeer, FromGoalsToActionCUI, baraglia2016initiative}.

Proactivity can be valuable in human–AI interaction. 
Studies show that proactive CAs can introduce new topics~\cite{proactiveagentwithInnerthoughts}, offer richer suggestions~\cite{ProactiveServiceBot}, and persuade users toward target behaviors~\cite{ProactivityWithANewNonCoopUser}. 
One domain where proactive CAs have been employed is supporting Knowledge Workers' (KW) productivity and well-being~\cite{KimaniAMBER, workFitCUI, KocielnikROBOTA, WilliamsDetachment, PITCHAdnan}. 
In these studies, CAs take the role of a productivity or well-being coach and regularly prompt KWs to articulate plans or reflect on their day, proactively leading the conversation. 
By initiating daily check-ins (e.g., morning and evening), CAs can help make users' intentions more concrete and actionable, thereby improving adherence to planning~\cite{LuoTandemTrack, KimaniAMBER, GroverTwoAgents}. 
Through reflective prompts, these agents can increase users’ awareness of everyday habits and foster behavior change~\cite{MolsReflection, KocielnikROBOTA, HoReflectionWithAgent}.  
These studies report positive outcomes such as greater engagement with work, improved work–life balance, better physical health, and more structured planning~\cite{KocielnikROBOTA, GroverTwoAgents, KimaniAMBER, workFitCUI, AhirePhysical, PITCHAdnan}.

\revision{However, we still lack an understanding of \emph{how} people actually communicate with CAs in everyday work, including what happens when the agent’s proactivity does not land well. 
This limits our ability to explain the mechanisms through which CAs support (or disrupt) desirable behavioral changes in the context of productivity and well-being.
These questions are especially important for LLM-based agents, which use human-like dialogue that encourages users to treat them socially~\cite{ComPeer, GPTCoach}, even though their suggestions can still be unpredictable—helpful at times, but also hallucinated, or poorly timed~\cite{rapp2025people, ComPeer}.
Recent work further shows that LLMs are poor at initiating grounding, leading to communication breakdowns~\cite{shaikh2024grounding, shaikh2025navigating}. 
Unlike task-oriented systems where breakdowns occur when the agent misunderstands user intent and fails to complete a task~\cite{myersHowUsersRepairVUI, resilientchatbots}, we anticipate additional forms of breakdown in proactive settings, where the agent’s behavior do not meaningfully support its overarching goal.
}

\revision{
To address these gaps, we examine how users engage with a proactive coaching agent and the deficiencies in agent behaviors that contribute to breakdowns. 
In our study, we define breakdowns as observable moments in the interaction where the participants and CA failed to maintain a shared understanding and conversational pace, leading to disruptions in the exchange~\cite{clark1991grounding, resilientchatbots, myersHowUsersRepairVUI, LeeMinGracefullyMitigatingBreakdowns}. 
For example, such cases can include moments when the conversation stalls, becomes unhelpful or dismissive, or otherwise diverges from the initial goal. 
In this paper, we explore this broader conceptualization of breakdowns in the context of proactive agent–user interactions based on observable behavior in conversations.
Understanding these breakdowns is critical for designing proactive systems that can improve conversational experience and foster cooperation in human-AI partnerships. 

To study these dynamics, we deployed a proactive, coaching-style agent (\sys{}) for two weeks with 12 graduate students. 
\sys{} checked in twice per day, once in the morning to facilitate planning and once in the evening to support self-reflection, resulting in 336 conversations (3{,}181 turns). 
We applied a codebook-based thematic analysis to generate qualitative patterns in planning, reflection, cooperation, and breakdown, and a user behavior dialogue-act analysis to quantify how often different behaviors occurred and how they varied across sessions and conditions.}
Specifically, we ask: 

\begin{itemize}
    \item \textbf{RQ1}: How do users engage with a proactive agent that facilitates daily planning and reflection?
    \item \textbf{RQ2}: What conversational behaviors of the coaching agent contribute to breakdowns in conversational interaction?
\end{itemize}

\revision{
At a high level, our analysis shows that users actively shaped the interaction: they adopted suggestions when useful, corrected the agent's assumptions, and used the chat to self-reflect on past activities and recognize desired behavior changes.
We also observed systematic breakdowns in the interaction, such as premature turn-taking and rigidity, which point to design opportunities for more flexible and context-aware proactive agents.

This study makes three contributions: (i) a thematic and dialogue-act analysis that provides an in-situ account of how users communicate with a proactive coaching agent; (ii) an anonymized dataset of 336 conversations (3{,}181 turns); and (iii) design considerations for building more adaptive and context-aware proactive CAs.
}

\section{Related Work}

In this section, we review prior work on conversational agents (CAs) that support behavior change, with a focus on how these systems shape users' planning, reflection, and everyday work practices.
We organize our review around a conceptual framework that guides our study, which consists of four dimensions: (1) planning and reflection, which describe how conversations help users articulate and evaluate their daily activities; (2) social interpretation of the agent, which influences how users perceive and respond to CA behaviors; (3) cooperation, which captures how users and agents work to maintain shared understanding during interaction; and (4) proactivity, which concerns how agents initiate and guide conversations rather than waiting for user input.
\revisionCHI{
Although these dimensions emphasize different aspects of conversational interaction, prior work suggests that they interact in shaping how users engage with CAs that support behavior change~\cite{chaves2021should, botOnSpeakingTerms}.
Because conversations leverage social intelligence, understanding users’ planning and reflection behaviors requires examining how they interpret the agent as a social entity.
Similarly, a CA's proactivity structures when and how opportunities for planning and reflection arise, for instance, through check-ins, follow-ups, or suggested next steps, but its effects depend on users' willingness to cooperate.  %((4)-(1)-(2)-(4)
In productivity contexts, such proactivity may also introduce a sense of accountability by positioning the agent as an ongoing social presence, which prior work suggests can support task tracking and adherence~\cite{PITCHAdnan, KimaniAMBER}.
Conversely, mismatches across these dimensions, such as misaligned interventions, unclear conversational moves, or users' uncertainties about system capabilities, can hinder meaningful interaction and lead to breakdowns~\cite{transparent-agents, MitchellAIMethodsForMicroCoaching, sajidJustStop}. 
These dimensions provide a lens for examining both how users engage with proactive planning agents and how specific conversational behaviors may hinder meaningful interaction. 
% it is important to understand how users treat the agent socially to study their planning and reflection behaviors.
% agent-initiated actions are socially interpreted and taken up 
% studying planning and reflection behaviors requires us to understand 
% Engagement with a proactive agent is not determined solely by the content of planning prompts, but by . %((1)/(2)/(3)).
% , 

}
% These dimensions are drawn from prior literature and motivate the design of our system.

\subsection{Planning and Reflection with Conversational Agents at Work}

HCI researchers have explored the role of conversational interaction to facilitate planning and reflection for KWs' productivity.
Several works developed an agent to collect self-reported tasks from users, manage scheduling, and facilitate work planning~\cite{KimaniAMBER, GroverTwoAgents, KocielnikROBOTA, MARIKYAN2022572}.
For example, Amber~\cite{KimaniAMBER} initiated conversations to elicit user's tasks, provided reminders for task scheduling, suggested breaks, and facilitated work reflections based on user interactions and context sensing. 
% Similarly, Robota was developed as a multimodal agent, combining voice and chat-based interactions to help users plan tasks, reflect on their motivation and satisfaction, and assess their learning~\cite{KocielnikROBOTA}.
At the same time, assistants also helped with automating some of the KWs' tasks, such as scheduling meetings~\cite{calendar.help}, managing teamwork~\cite {TaskBotPaper}, finding work resources~\cite{LiaoWorkPlaceSociality}, organizing email inboxes~\cite{emailRADARTaskmanagement}, and managing to-do lists~\cite{todolistagent}. 
Findings from these works highlight benefits for productivity such as increased plan adherence and deeper work engagement~\cite{KimaniAMBER, KocielnikROBOTA}.

Beyond productivity, CAs have also been studied to promote well-being. 
This includes prompting healthy breaks~\cite{workFitCUI, CamboPhysical, AhirePhysical}, facilitating reflection on work-life balance~\cite{WilliamsDetachment}, and helping to manage stress~\cite{meyer2019enabling, folstad2019different}. 
Similar works have explored how KWs can be encouraged to take work breaks for their physical and mental well-being to mitigate the negative effects of sedentary lifestyle~\cite{AhirePhysical, CamboPhysical, workFitCUI}.
Research shows that engaging in reflective conversations can facilitate self-awareness, helping people be more aware of task switching, track progress, evaluate obstacles, and reflect on higher-level goals~\cite{PlanningHabit, StudyBuddyPaper, MARIKYAN2022572, KocielnikROBOTA, KimaniAMBER, giveAndTake}. 
Such reflection fostered healthier work practices and greater satisfaction, particularly in fast-paced, busy environments where self-reflection is often overlooked~\cite{KocielnikROBOTA}.

\revision{
Although the above works show that users improve their productivity and well-being by interacting with CAs, most studies have focused on outcomes (e.g., productivity, stress reduction, perceived usefulness) or user perceptions, but have not closely observed users' conversational behaviors. 
Analyzing these behaviors provides a deeper understanding of human-ai interaction by identifying recurring patterns, assessing user engagement, and evaluating the effectiveness of conversational design~\cite{CasasTrendsofChatbotEval}.

Moreover, prior studies were conducted in the era before LLMs~\cite{KocielnikROBOTA, KimaniAMBER, WilliamsDetachment, GroverTwoAgents}, where the agent's role was more reactive to user input and limited because of the static, fixed dialogue structures. 
Such structural limitations undermines the quality of reflection as users do not get a chance to engage deeply in response to the posed questions~\cite{reflectionInTheoryandPractice}.
LLMs now enable more dynamic interactions by exhibiting better conversational intelligence~\cite{ConversationalAbility, sorin2024large}, persuasion~\cite{AIPersuasionVaccination, salvi2025conversational}, and novelty in interactions~\cite{CueingTheCrowd}. 

Building on these capabilities, recent works have begun exploring AI in a coaching role~\cite{LanceRhoCommCoach, CoachingCoPilot, GPTCoach, PlanFitting}, yet we still lack a detailed understanding of how users actually behave in dialogue with such coach-like agents---how they accept, negotiate, resist, or dismiss the agent’s suggestions during daily planning and reflection.
Our work addresses this gap by examining in-situ conversation logs from a fourteen-day longitudinal study, focusing on the emergent behaviors that characterize how people engage in daily planning and self-reflection with an LLM-based coach.
}

\revision{
\subsection{Social Perception of Conversational Agents}
\label{sec:social-relatedwork}
Users’ perceptions of a CA’s social presence strongly shape how they engage with it~\cite{botOnSpeakingTerms}.
Prior research consistently shows that when people interpret a CA as a social partner rather than a technical tool, they are more likely to reflect on their behaviors, sustain interaction, and incorporate suggested changes into their routines~\cite{chaves2021should, rapp2025people, CoachingCoPilot, LiaoWorkPlaceSociality}. 
In productivity contexts, for example, agents that follow up on users’ stated plans can evoke a sense of social accountability~\cite{PITCHAdnan, WicaksonoPlansDeliberate, MitchellAIMethodsForMicroCoaching}.
Because conversational interfaces mimic human dialogue, users often respond with politeness, humor, self-disclosure, or repair behaviors consistent with the CASA paradigm\footnote{The Computers Are Social Actors (CASA) paradigm argues that people apply social norms and expectations from human–human interaction to computers when those computers display social cues~\cite{nass1995can}.}~\cite{AlexaIsMyBFF, LiaoWorkPlaceSociality, SeeringBabyBot}.
Yet users differ in their willingness to treat agents as social entities: some readily maintain engagement and work to repair breakdowns, while others view the CA strictly as a tool and disengage after a single error~\cite{earlyquittersPaper, LeeMinGracefullyMitigatingBreakdowns}. Research on anthropomorphism similarly reports mixed effects, sometimes increasing satisfaction~\cite{AlexaIsMyBFF}, but in other cases, frustrating or even unsettling users~\cite{GroverTwoAgents, rapp2025people}.

% why we should study cooperation
Because users’ social perceptions shape their willingness to engage, persist, and repair, cooperation becomes the key interactive process through which a CA and user work toward shared goals~\cite{clark1991grounding}.} 
In task-oriented scenarios, where users interact with AI systems to achieve a concrete outcome (e.g., processing a return with a customer service bot or booking a ticket with an airline virtual agent), cooperation is critical to reaching the goal of the interaction~\cite{CollaboratingTextBasedAmonRapp, laban2019working}.
Prior work has shown that when conversational agents are perceived as cooperative, users consider their information richer, rate the interaction as higher quality, and report greater satisfaction~\cite{laban2019working, hoffman2004collaboration}. 
To cooperate, people often adapt their own behavior: simplifying or reformulating messages, rephrasing to repair breakdowns, or even using gestures with robots to accommodate system constraints~\cite{CollaboratingTextBasedAmonRapp, BankingChatbot, myersHowUsersRepairVUI, VUIEveryDayLifeJoelFischer, whyThatNaoPelikan}. 
However, this process is also challenging because chatbots often display limited intelligence while users bring expectations of human-like competence and conversational norms, leading to mismatches and frustration~\cite{BadPALugerSellen, realconvoswithchatbot}.
Unsatisfactory responses such as turn-taking failures, hallucinations, incorrect information, or lack of flexibility undermine user trust and can ultimately reduce engagement~\cite{earlyquittersPaper, resilientchatbots, AkhtarAnalysisChatbotConversation, dechant2025future}. 
\revision{
While we know how cooperation is enacted in task-centric chatbots~\cite{CollaboratingTextBasedAmonRapp}, we lack a clear account of how it unfolds when users plan and reflect with an LLM coach that provides productivity and well-being advice. }

\revision{
Our study fills this gap by examining how users perceived the agent as a social entity influencing their planning and reflection behaviors. 
User behaviors in this setting would be different than in task-centric scenarios because unlike service chatbots that follow a request–response–confirmation template, our system involves users in  relatively open-ended planning and reflection.
The interactions are less about completing a transaction and more about scaffolding self-reflection, prompting users to accept, decline, or negotiate suggestions. 
In this context, cooperation involves more than providing slot values--it includes sharing context, disagreeing or negotiating advice, and aligning conversations with one’s preferences. 
% We also observe how users treated the agent socially: 
% updating it on progress or providing additional context. 
% By analyzing these behaviors, we show how users manage cooperation with a coach-like agent and how their actions signal perceptions of the agent’s social presence.
}

\subsection{Designing Proactive Agents for Behavior Change}

% \cite{montenegro2019dialogue} The chatbot in this study is a virtual coach for elderly people. Its role is to coach elderly users to help them maintain a healthy, independent lifestyle.
% they used a Wizard of Oz (WoZ) setup:
% A human “wizard” played the role of the chatbot, using scripted behaviors consistent with the taxonomy.
% 192 elderly participants (across Spain, France, Norway) interacted with it.
A longstanding debate in HCI concerns how much initiative should rest with users versus systems. 
Shneiderman argued for direct manipulation interfaces where ``users comprehend the display, feel in control, [and] the system is predictable'' so they remain responsible for outcomes~\cite{ShneidermanMaes}.
Maes, by contrast, saw growing complexity as making delegation inevitable: software agents should be personalized, proactive, long-lived, and adaptive, taking initiative on behalf of overloaded users.
Framing these positions as a strict opposition has proven limiting~\cite{ShneidermanMaes}. 
Horvitz later described this as a “false dichotomy” and advanced principles of mixed-initiative interaction, where automated services and direct manipulation are coupled~\cite{HorovitzMixed}.  
\citeauthor{HorovitzMixed}'s Lookout system, for example, anticipated calendar events from email but always allowed users to confirm or refine results. 
Allen extended this view to dialogue, defining mixed-initiative as a flexible strategy in which initiative is dynamically negotiated: sometimes the system reports or clarifies, sometimes the user directs, and sometimes roles shift opportunistically~\cite{AllenMixedInitiative}.

Building on this lineage, recent work has asked whether conversational agents should be reactive or proactive.
Reactive agents respond only when the user initiates, giving users more control but limiting the system’s role. 
Proactive agents, by contrast, initiate interactions by offering reminders, or suggesting actions~\cite{ComPeer, FromGoalsToActionCUI, baraglia2016initiative}. 
\citeauthor{chaves2021should} describes proactivity as sharing initiative with the user, where the agent may start exchanges, propose topics, or offer follow-up questions. 
\revision{In the context of coaching, people want CAs to be proactive in making suggestions to improve their productivity~\cite{ProductivityTelemetryAgent, chaves2021should, lee2025choir}. 
Research shows that agent proactivity can introduce new topics~\cite{proactiveagentwithInnerthoughts}, provide richer suggestions~\cite{ProactiveServiceBot}, and persuade users toward a target behavior~\cite{ProactivityWithANewNonCoopUser}. }
Researchers have also mapped proactivity on a spectrum to show how much initiative rests with the user versus the system. 
\citeauthor{isbellLevelsOfProactivity} outlines an Interface-Proactivity continuum that ranges from the user doing everything on their own to the system acting on their behalf, with several mixed approaches in between where the system may highlight information or suggest options.  
Recent efforts to push AI toward agentic behavior, such as the release of the ChatGPT agent~\cite{openai2025chatgptagent}, illustrate the far end of this spectrum where the system can act on the user’s behalf. 
Yet, highly proactive designs can also be perceived as intrusive, disruptive, or controlling~\cite{LiaoWorkPlaceSociality, ProactiveServiceBot, tallynEthnobot}.
With the rise of LLM-powered agents, the model responses can exhibit agency in unpredictable ways that produce negative emotional effects toward the technology~\cite{rapp2025people}. 
% This tension frames the ongoing debate about how initiative should be shared between user and system.
% and challenges around timing and intrusiveness, with users often perceiving proactive interventions as annoying or misaligned with their needs~\cite{calendar.help, ContextAwarenessIssuesBellotti}. 

Across these strands of work, the challenge is balancing initiative in ways that adapt to users’ evolving needs. 
Proactivity can keep conversations goal-oriented~\cite{ProactivityWithANewNonCoopUser}, but it can also constrain flexibility when users’ contexts shift. 
Our study extends this literature by examining how users engage with an LLM-based coach that proactively decides planning and reflection goals, highlighting how proactivity can be helpful and when it can go out of context.

\revisionCHI{}
% \revision{
% \subsection{Conceptual Framework}

% The four areas we outlined from the review above: planning and reflection, social perception of the agent, cooperation, and proactivity are inter-related concepts. 
% Since conversations leverage social intelligence, studying planning and reflection behaviors requires us to understand how users treat the agent socially. 
% In our study, users cooperated with the agent by reading and answering its questions, elaborating on their activities or feelings, and taking up its suggestions, opening space for planning and reflection. 
% This cooperation was supported by the agent’s proactivity as it checked in, asked reflective questions, and proposed next steps. 
%  Such proactivity is meaningful in productivity and well-being contexts because the agent actively elicits, shapes, and extends the user’s plan as opposed to reactive tools that require users to initiate planning (e.g., journals or to-do lists ).
% A proactive agent, by contrast, participates in co-planning by externalizing questions users may not ask on their own.
% Studying how users engage with this proactive, conversational agent therefore helps explain how planning and reflection emerge through a blend of social interpretation, cooperation, and system-initiated guidance.
% }

\begin{figure*}[t!]
    \centering
    % subfigure [t] + \vtop is the gold standard for top alignment
    \begin{subfigure}[t]{0.48\textwidth}
        \centering
        \vtop{\vskip0pt\hbox{\includegraphics[width=\linewidth]{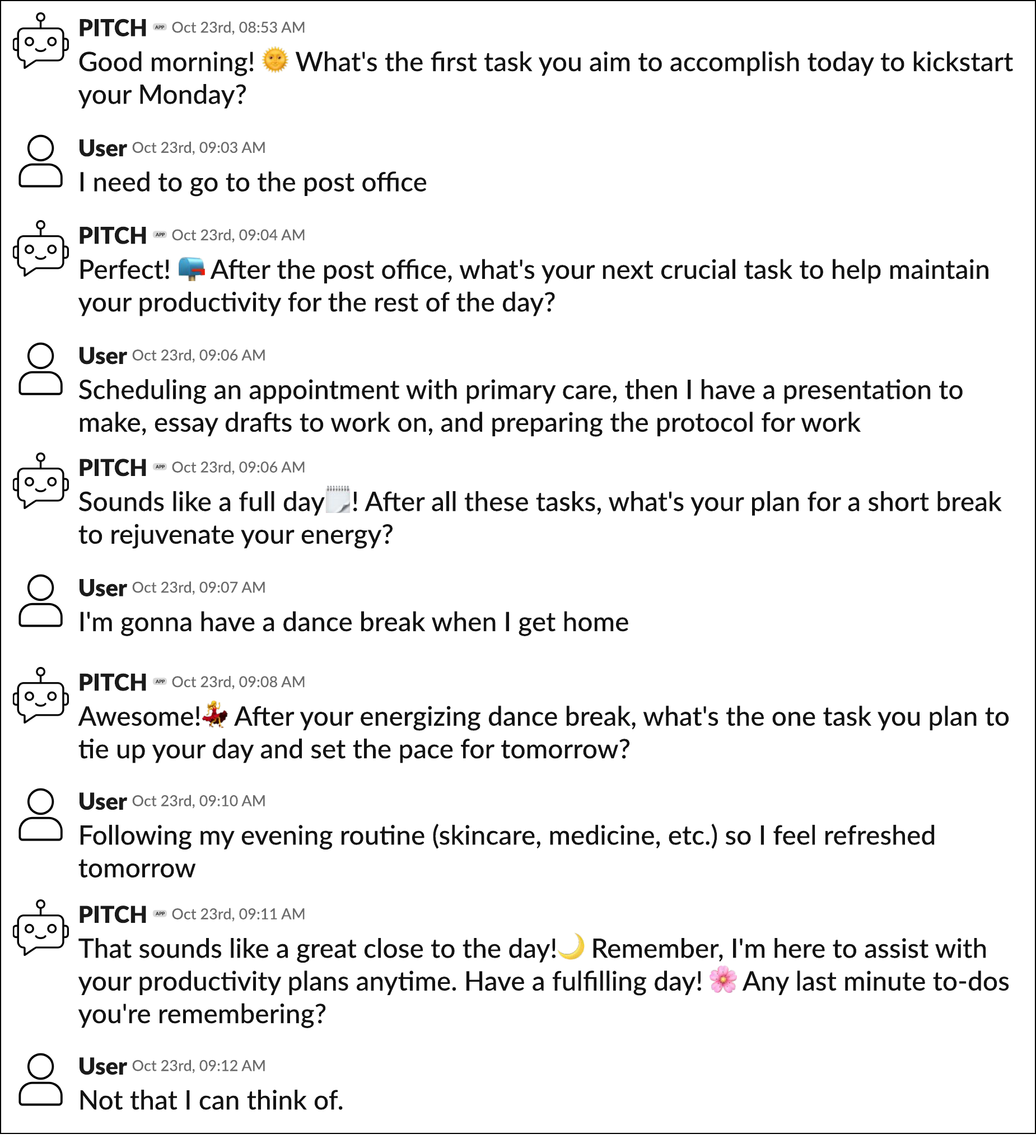}}}
        \caption{Morning Plan Externalization}
        \label{fig:morningconversation}
    \end{subfigure}
    \hfill
    \begin{subfigure}[t]{0.48\textwidth}
        \centering
        \vtop{\vskip0pt\hbox{\includegraphics[width=\linewidth]{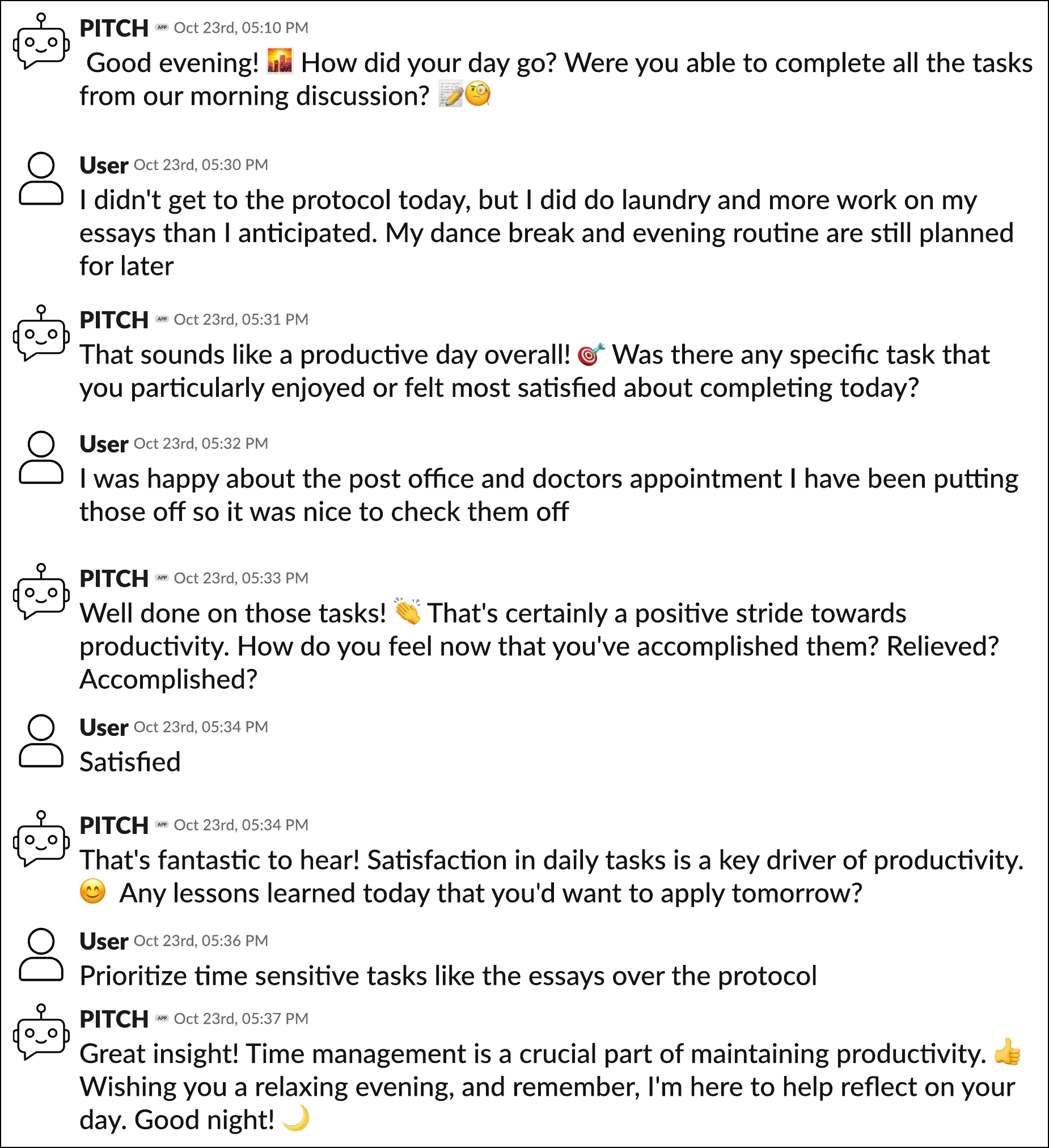}}}
        \caption{Evening Reflection}
        \label{fig:eveningconversation}
    \end{subfigure}
    \caption{An example of the morning and evening conversations from the field study. \sys{} checks in with the user in the morning to externalize their plan (a) and follows up in the evening to support self-reflection (b). }

    \label{fig:Scenario1}
\end{figure*}

\revisionCHI{
\section{\sys{} - A
Proactive Agent for Daily Planning and Self-Reflection}
\subsection{Design Goals and Implementation}

We designed two versions of \sys{} that participants used each week in random order: a fixed-goal version (\sys-N) and a rotation-based version (\sys-R). 
We designed \sys{} to support reflection in users’ plans, with morning conversations prompting them to externalize their intentions and evening conversations returning to the externalized plan to support self-reflection. 
Below, we detail the design specifications.

\subsubsection{Daily Conversational Interaction for Morning Plan Externalization}
\label{sec:morning}

\sys{} was designed to act as a productivity coach that proactively checks in with a worker each morning. The morning interaction centered on planning, with the goal ``to help workers structure their day to enhance productivity and well-being (\sys{}-N).''
By answering \sys{}’s questions, users can be reminded of the tasks they consider important for the day, and the conversational format prompts people to articulate task details they may not otherwise specify in a traditional task list (e.g., priorities, timing, and expected effort). 
Prior work shows that externally articulating intentions supports prospective memory~\cite{gilbert2015strategic}, and mentally simulating a task can improve plan adherence~\cite{doi:10.1080/17470218.2014.956127}.

The morning conversation therefore serves as the point of ``externalization'' that the evening interaction later reflects back on. 
This connection to users' day’s plans is a key design choice for studying self-reflection: participants first lay out a concrete, self-generated plan, which then becomes the shared reference point for evaluating their day through conversation.
Example prompts used in the morning included:

\begin{itemize}[leftmargin=5.5mm]
    \item Good morning! What are the top three tasks you plan to accomplish today to make your Wednesday productive?
    \item Good morning! Since it's Saturday, what enjoyable activity have you planned for yourself today?
\end{itemize}

\noindent  The full morning prompt is available in Appendix~\ref{app:morningprompt}. An example morning interaction is shown in Figure~\ref{fig:morningconversation}.
% This balance between productivity and well-being is also reflected in the rotation strategy (Section~\ref{sec:rotationdev}). 

\subsubsection{Personalized Reflections on Morning Plans}
\label{sec:evening}

To support self-reflection in the evening, we leveraged the morning conversation history to enable \sys{} to ask questions grounded in what the user had planned earlier that day. 
In the evening system prompt, the assistant’s goal was defined as ``to help workers reflect on their day for improved productivity and well-being, (\sys{}-N)'' and the prompt explicitly instructed the agent to keep questions ``within the context of the morning conversation.''

Including the full morning context operationalizes self-reflection in a concrete way: users are asked to revisit the intentions they articulated earlier, compare them with what actually happened, and describe reasons for fulfillment or deviation. 
This allowed us to observe reflection behaviors because participants are prompted to evaluate their plans, reconstruct their day, and make sense of successes or difficulties. 
Prior work has used prior conversational context to guide empathetic responses~\cite{seo2024chacha}, journaling~\cite{DiaryMate}, and well-being monitoring~\cite{LLMsForPublicHealth}. 
Here, we used this technique to ensure that reflective questions are not generic but tied to the user’s own earlier commitments. 
In contrast, some previous approaches relied on general-purpose questions such as, “How much progress do you feel you made on your scheduled task?”~\cite{KimaniAMBER}.
Example evening prompts included:

\begin{itemize}[leftmargin=5.5mm]
\item Good evening! Hope your day went well. Were you able to complete the main paragraph of your lit review as planned?
\item Good evening! How was your day practicing mindfulness through body scans? Did you find any satisfying moments? 
\end{itemize}

\noindent Each question directly referenced the morning plan, making the reflective behavior observable in the dialogue.
The full evening prompt is available in Appendix~\ref{app:eveningprompt}. 
An example evening conversation is shown in Figure~\ref{fig:eveningconversation}.

\subsubsection{Lightweight Check-ins for Sustained Engagement}

We ensured that \sys{} stayed focused on its conversational goal without becoming formal or excessively verbose, since brief check-ins are more suitable for long-term engagement~\cite{MitchellAIMethodsForMicroCoaching}. 
Early pilot tests showed that \sys{} sometimes extended conversations through multiple follow-up questions and its verbosity. 
While extended conversations can be beneficial, our aim was to support the users' thought process through drawing questions rather than \sys{} conveying information. 
Moreover, earlier studies showed that KWs do not prefer extended conversations as they are short on time~\cite{GroverTwoAgents, ProductivityTelemetryAgent}.
This included prompting the model to keep responses short and easy to understand, an observer agent checking for conversation length, using emojis appropriately, and enforcing a 30-word limit per message.

\subsubsection{\sys-R : Rotation Strategy}
\label{sec:rotationdev}
\sys{} also explored whether dynamically rotating goals each day affected engagement and introduced variation in users' reflection behaviors. 
The default planning and reflection goals described in Sections~\ref{sec:morning} and \ref{sec:evening} were used for \sys{}-N. 
In contrast, \sys{}-R used more specific goals drawn from self-help literature and academic work. 
To reduce redundancy, the order of goals was shuffled.
A randomly chosen morning goal and a randomly chosen evening goal (from $N$ possible goals each) produce $N^2$ possible daily combinations. 
This design increases diversity in conversation topics and creates opportunities for studying how conversational behaviors vary with different productivity and well-being topics.
% We included seven morning and seven evening goals, with three focused on well-being in each. 
The detailed system architecture and conversation goals are described in our previous work~\cite{PITCHAdnan}.
% We deployed two versions of the CA:

% \begin{enumerate}
%     \item \textbf{\sys{}-R} (rotation), which cycled through different goals each day.
%     \item \textbf{\sys{}-N} (no rotation), which kept the same goal every day.
% \end{enumerate}

\subsubsection{Technical Implementation}
We implemented \sys{} as a chatbot application in Slack~\cite{slackUnlockYour}, a platform widely used by KWs, so that users would not need to leave their existing work environment to interact with the system.
We built \sys{} using the Slack Bolt framework for Node.js.
For message generation, we used OpenAI’s GPT-4 API.
MongoDB was used to store daily conversations and associated metadata.
Further technical details can be found in our previous work~\cite{PITCHAdnan}.
}
\section{Method: Field Deployment Study}

In October 2024, we conducted a two-week (14 weekdays) deployment study to investigate how users interact with \sys{}, which initiated morning conversations to help externalize their plans and evening conversations to support self-reflection. 
The Institutional Review Board (IRB) of the researcher’s university approved the procedure of our study, and informed consent was obtained from all participants. 
We organized the study procedure into three main phases: an onboarding session, a field deployment, and a post-study interview session.
While we conducted post-study interviews and collected diary survey responses, we published the results that focused on understanding users' perceptions and experience~\cite{PITCHAdnan}. 
In this study, we focused on conversation analysis from the same study to understand how participants behaved with \sys{}, and we did not include any results from the interview or survey data. 

\subsection{Study Procedure}
We conducted onboarding sessions in person or over Zoom to introduce participants to \sys{}. 
During onboarding, we demonstrated the Slack-based chatbot, explained the study procedure, and provided written instructions \revision{provided in Appendix \ref{appendix:written-instructions}.}
Participants were invited to a dedicated Slack workspace and asked to complete a pre-study demographic form (age, gender, race/ethnicity, occupation, daily activities, and time spent). 
Following onboarding, participants began a two-week field deployment. 
Each day, \sys{} initiated check-ins at user-specified times in the morning and evening. 
Participants engaged with these prompts and completed a brief daily survey. 
All interactions with \sys{} were conducted in English. 
We collected all conversation interactions between participants and \sys{} during the study.  
At the end of the two-week study, we did a semi-structured debriefing interview. 
Participants received \$50 as compensation for their participation. 
  
\subsection{Participants}

\revision{We advertised the study through broad university channels, including institution-wide bulletin boards and targeted mailing lists for graduate students in graduate school, as well as the author's department.
We recruited 12 participants who met our inclusion criteria: (1) were over 18 years old, (2) were fluent in verbal and written communication in English, (3) were familiar with Slack, (4) were interested in interacting with a productivity AI agent, and (5) were managing different types of daily tasks. 
All participants were full-time graduate students (2 MS, 10 PhD) in Computer Science, with one exception (Physics), at the researchers' university, and had a basic knowledge of computing.
% ~\cite{374}.
% We recruited participants with varying levels of time management skills to capture a range of user-agent interaction strategies (e.g., structured vs. ad-hoc planners).
Graduate students in STEM fields, although not representative of all knowledge workers, face many demanding characteristics that knowledge workers have, including independent task management, competing deadlines, flexible working hours, various responsibilities, and blurred work–life boundaries.
This made them an appropriate exemplar group for an initial \revisionCHI{\textit{exploratory}} investigation into planning and reflection interactions with proactive conversational agents, whose demands are amplified compared to those of average knowledge workers.
As part of the pre-study questionnaire, participants self-reported their frequency of Slack use and their time-management skill level by answering multiple-choice questions. 
These measures were collected only to describe the sample and are not used as variables in our analysis.}
Of the 12 participants, 6 identified as male, 1 as non-binary, and 5 as female. 
Their ages ranged from 22 to 35 ($\mu$ = 26.91; $\sigma$ = 3.87). 
\revisionCHI{Slack use frequency also varied across participants: 33.3\% reported always using Slack, 33.3\% often, and 33.3\% sometimes.}
Demographic information is summarized in Table~\ref{tab:demographics}.

% Our objective was not to estimate population parameters but to derive a behavioral understanding of how people interact with a proactive planning agent embedded in their existing workflows. Thus, a graduate-student sample—coherent in context yet diverse in daily responsibilities—was methodologically appropriate for producing the type of fine-grained conversational analysis we undertook.

% therefore we recruited people who had varying levels of time management skills and 
% HCI - 8 participants
% ML/AI research - 2 participants 
% Physics - 1
% Recruited participants were majoring in fields HCI, CS, and Physics and had basic knowledge of computing.
\begin{table}[htbp]
\caption{Participant demographics and self-reported metrics. Gender abbreviations: M = Man, W = Woman, N-B = Non-binary. Participants managed diverse tasks including research, coursework, teaching, and service responsibilities with varying time allocations.}
\centering
\small
\begin{tabular}{p{0.06\linewidth} p{0.12\linewidth} p{0.10\linewidth} p{0.10\linewidth} p{0.40\linewidth}}
\toprule
\textbf{PID} & \textbf{Gender / Age} & \textbf{Race} & \textbf{Time Mgmt} & \revision{\textbf{Typical Daily Activities}} \\
\midrule
P1  & M / 25  & Asian & Fair & Research \textit{(2--3h)} || Coursework \textit{(7h/wk)} || Meetings \textit{(3--4h/wk)} \\
\midrule
P2  & M / 33  & White & Good & Research \textit{(3--9h)} || Teaching \textit{(1.5h Tu/Th/Fr)} || Meetings \textit{(1h/wk)} || Programming \textit{(variable)} \\
\midrule
P3  & M / 24  & Asian & Good & Research \textit{(5h)} || Coursework \textit{(3h)} || Programming \textit{(3h)} \\
\midrule
P4  & W / 25  & Asian & Good & Teaching/Grading \textit{(1h)} || Research \textit{(4h)} || Coursework \textit{(4h)} \\
\midrule
P5  & M / 27  & Asian & Good & Research \textit{(variable)} || Office Hours || Coursework \textit{(2h)} \\
\midrule
P6  & W / 27  & Asian & Fair & Research || Coursework || Meetings || Coding || Classes \\
\midrule
P7  & W / 35  & White & Fair & Teaching \textit{(2h)} || Meetings \textit{(5--6h)} || Programming \textit{(3h)} \\
\midrule
P8  & W / 30  & Asian & Fair & Research \textit{(4--5h: programming, writing, reading)} || Meetings \textit{(2--3h)} || Mentoring \textit{(1h)} \\
\midrule
P9  & M / 25  & White & Fair & Teaching \textit{(1h)} || Meetings \textit{(0.5h)} || Coursework \textit{(1h)} || Research \textit{(2h)} \\
\midrule
P10 & W / 24  & Asian & Very Good & Class \textit{(5h)} || Work Shift \textit{(5h)} || Coursework \textit{(5h)} || Other \textit{(3h)} \\
\midrule
P11 & N-B / 22 & White & Fair & Research \textit{(4h)} || Classes \textit{(2--4h)} || Work Meetings \textit{(0--1h)} || Club Meetings \textit{(0--1h)} \\
\midrule
P12 & M / 27  & Asian & Fair & Research || Meetings || Programming || Writing \\
\bottomrule
\end{tabular}
\label{tab:demographics}
\end{table}

\subsection{Data Analysis}

\subsubsection{Codebook-based Thematic Analysis}
% what we did and why we did it
\revision{
We conducted a codebook-based thematic analysis to surface and articulate semantic behavioral patterns in the data~\cite{clarke2015thematic, braun2021oneSizeFitsAll, braun2006using}. 
The analysis was primarily inductive, where codes and themes were developed through close engagement with the data rather than from pre-existing literature or a framework, while remaining theoretically informed by our research aims (understanding user behaviors in daily planning and reflection). 
Our epistemological stance was informed by  essentialist/realist approach where we treat participants’ conversations as meaningful access to experience, while acknowledging the role of context and interpretation~\cite{braun2006using}.
% Our research values and philosophical orientation were informed by an essentialist/realist approach where the underlying assumption is that participants’ talk provide meaningful access to experience, while acknowledging the role of context and interpretation
% language reflects and enables us to articulate meaning and experience~\cite{braun2006using}. 
% We sought to develop themes as patterns of behaviors in the data and to translate these into actionable design implications, reflecting the applied and pragmatic orientation typical of codebook-based analysis. 

% how we did: procedure details
The conversation data was stored in a spreadsheet where each row was the user's or chatbot's turn. 
Data columns were: participant ID, date, day, session (Morning/Evening), timestamp, and response text.  
% six phases outlined by braun and clarke
The first author familiarized themselves with the data, taking notes, and
then generated initial codes systematically across the dataset (e.g., ``user shared a ranked task list'' or ``agent suggested user to break down their task''), documenting them in a separate spreadsheet with code IDs and descriptions. 
Each response by the agent or user was considered in the analysis and was tagged with either one code, multiple codes, or no codes, as relevant. 
After coding roughly 50\% of the data, 226 codes were generated. 
To refine and stabilize the codebook, the first author and three other researchers met to collectively review and discuss the codes with the aim to merge repetitive codes and clarify ambiguous ones.
When any code was unclear to the team, we referred back to the conversation transcripts to ground the code's meaning within its specific conversational context.
This collaborative session established shared understanding and consistency in code definitions.

Using the refined codebook, the first author then coded the remaining 50\% of the data, resulting in a total of 439 codes.
Thematic saturation was achieved during the final coding cycles, as the last two participants' conversations yielded no new codes.
Next, the first author and three other researchers met to search for candidate themes by collating and grouping related codes.
The 439 codes were transferred from the spreadsheet to a shared Miro board~\cite{miro} as sticky notes to facilitate collaborative theme development and visualization through affinity diagramming.
The team created initial theme headings and systematically analyzed each code, placing it under the most appropriate theme.
When codes did not fit cohesively within existing themes, new themes were proposed and discussed among the researchers.
This process resulted in the identification of 32 candidate themes, organized into two analytical categories: 13 themes capturing behavioral patterns in the chatbot's responses and 19 themes capturing patterns in users' behaviors (Tables~\ref{tab:user-themes}, \ref{tab:bot-themes} in Appendix~\ref{sec:appendix-data-analysis}).
The process of refining themes was iterative, involving four in-person research team meetings to review the visual board, assess internal coherence, and ensure distinctions between themes.
Throughout this process, codes were merged, revised, or repositioned as needed.
% Throughout this process, themes were reviewed against both the coded extracts and the full dataset to ensure adequate representation and grounding in the data.
}

% In this round, some of the codes were merged. 
% With the shared understanding, 
% Then the first author again went back to code the remaining data and the 3 other researchers met to map the codes to the existing themes or create new themes. This process was iterative and was conducted across four sessions. 
% This iterative, collaborative process ensured that the final themes captured recurring interactive behaviors across the dataset.
% grouped codes into preliminary themes and merged overlapping ones. 
% Drawing on this shared framework, the first author coded the remaining dataset while iteratively refining the codes. 
% Once the full dataset was coded, the team held additional sessions to review the coding, map codes to themes, and refine the thematic structure through discussion. 

\subsubsection{User Behavior Dialogue Act Analysis}
\label{sec:dialogueact-analysis}
\revision{
To complement the thematic analysis and examine the frequency of user behaviors, we conducted a dialogue act analysis. 
Dialogue act analysis is a method from linguistics and NLP that examines each utterance in conversations to understand the conversational and linguistic behaviors speakers engage in~\cite{traum200020DialogueActTaxonomy, jurafsky1997switchboard, alexandersson1997verbmobil, PersuasionForGood}. 
For instance, for an utterance ``\textit{I am sorry},'' a dialogue act can be \emph{Apology}~\cite{jurafsky1997switchboard}.
Researchers have designed different taxonomies for conversations in unique contexts (e.g. task-oriented, casual) as the nature of dialogue in every context differs, resulting in different frequency of dialogue acts~\cite{traum200020DialogueActTaxonomy}.

Similar to prior works~\cite{wei2023leveraging, TestingBotIdentityAndInquiry, PersuasionForGood} in human-AI interaction, we also developed and customized our dialogue act codebook by adapting established taxonomies~\cite{traum200020DialogueActTaxonomy, jurafsky1997switchboard, alexandersson1997verbmobil} and extending them with acts grounded in our thematic analysis. 
From established taxonomies~\cite{ jurafsky1997switchboard, alexandersson1997verbmobil, montenegro2019dialogue} , we retained acts that were relevant to our domain (e.g., \emph{Comply}, \emph{Refuse}, \emph{Ask}, \emph{Contextualize}, \emph{Ignore}, \emph{Thank/Greet}), which capture fundamental communicative functions in dialogue. 
We additionally created five acts tailored to our specific context of planning and self-reflection support: \emph{Plan}, \emph{Report}, \emph{Self-reflect}, \emph{Recognize}, \emph{Change}, and \emph{Belated}. 
These domain-specific acts emerged from our thematic analysis and reflect the core behaviors our agent was designed to support. 
All 17 dialogue act categories, along with their descriptions, references, and examples, are provided in Table~\ref{tab:user-behaviors}.

To establish reliable coding, we employed deductive coding where two researchers independently coded 40\% of the data using the preliminary codebook. 
Discrepancies were resolved through discussion, refining category boundaries. 
Inter-rater reliability was assessed using Cohen's Kappa, with all individual category values exceeding 0.8 and overall agreement at $\kappa$ = 0.90, indicating strong agreement (for per-category $\kappa$ values, refer to Table~\ref{tab:irr-per-category} in Appendix~\ref{sec:appendix-data-analysis}). 
The first author then applied deductive coding by systematically assigning the finalized dialogue act codebook to the remaining 60\% of the data. 
Multiple dialogue acts could be assigned to a single utterance to capture overlapping communicative functions (e.g., \textit{``Quick meditation is a great idea. I will try today''} was coded as both \emph{Commend} and \emph{Comply}).
}
% There are a number of different dialogue activities that people have been designing taxonomies of dialogue acts for. Some examples include casual conversation (Jurafsky et al., 1997),
% classroom discourse (Sinclair and Coulthard, 1975), and various flavors of task-oriented dialogue, such as information seeking (van Vark et al., 1996), collaborative scheduling (Alexandersson et al., 1998), and direction following (Carletta et al., 1997).

\subsubsection{Quantitative Analysis}
\label{quant-analysis}
\revisionCHI{
We used generalized linear mixed-effects models (GLMMs) with a Binomial distribution and a logit link function in R (package \texttt{lme4})~\cite{bates2015fitting} to analyze how dialogue act frequencies varied across sessions and conditions. 
% We adopted a GLMM approach because the dependent variable consist of binomially distributed count data (the number of utterances labeled with a given dialogue act out of the total number of utterances per session).
We adopted a GLMM approach because we predict a binomially distributed variable, i.e., the proportion of messages labeled with a given dialogue act (e.g., Plan) out of the total number of messages in a conversation.
% For each dialogue act type, we fit binomial GLMMs where the dependent variable was specified as the count of utterances labeled with that act relative to the total number of utterances in that session. 
The model included three fixed effects: Time (Morning vs. Evening), Condition (Fixed vs. Rotation), and Day index (Z-score standardized (mean = 0, SD = 1)). Participant was included as a random intercept to account for repeated measures within participants.

The linear predictor for the expected log-odds of a proportion of a particular dialogue act for observation $i$ from participant $j$ can be expressed as:

\begin{align}
\eta_{ij} &= \beta_0 + \beta_1 \cdot \text{Time}_{ij} + \beta_2 \cdot \text{Condition}_{ij} + \beta_3 \cdot \text{Day}_{ij} + u_{0j}
\end{align}

where the proportion of the dialogue act is $p_{ij} = \text{logit}^{-1}(\eta_{ij})$, and the count of dialogue acts follows a binomial distribution with probability $p_{ij}$ and denominator given by the total number of utterances.

The parameters are defined as follows:
\begin{itemize}
  \item $\eta_{ij}$: linear predictor (log-odds) for observation $i$ from participant $j$.
  \item $\beta_0$: fixed intercept (baseline log-odds for the reference category).
  \item $\beta_1, \beta_2, \beta_3$: fixed-effect coefficients for Time, Condition, and Day, respectively.
  \item $\text{Time}_{ij}$: binary indicator ($1$ = Morning, $0$ = Evening).
  \item $\text{Condition}_{ij}$: binary indicator ($1$ = Rotation, $0$ = Fixed).
  \item $\text{Day}_{ij}$: Z-score standardized day index (mean = 0, SD = 1).
  \item $u_{0j}$: random intercept for participant $j$, accounting for repeated measures and between-participant variability.
  \item $u_{0j} \sim \mathcal{N}(0, \sigma_u^2)$
\end{itemize}

% Day was scaled using Z-score standardization to ensure comparability across models and to aid model convergence. 
% For models exhibiting convergence issues (REFUSE and SELFDISCLOSE), we used the BOBYQA optimizer with increased iteration limits (\texttt{optCtrl = list(maxfun = 1e5)}).
All models were assessed for overdispersion using the performance package~\cite{ludecke2021performance} to verify that the binomial assumption was appropriate for the data.
No significant overdispersion was detected for any dialogue act ($p > .05$), confirming the suitability of the binomial distribution.
Significance testing was conducted using the summary output from fitted models, which provides $z$-statistics and $p$-values based on the Wald test approximation for fixed effects in GLMMs.

}

\revision{Additionally, we analyzed the emotional tone of user messages using sentiment analysis. We computed sentiment scores for each user utterance using the RoBERTa model from TweetEval\revisionCHI{~\cite{barbieri-etal-2020-tweeteval}}, which assigns scores ranging from -1 (negative) to +1 (positive) by subtracting the probability of negative sentiment from positive sentiment. 
We then aggregated sentiment scores at the session level (averaging sentiment across all turns in a session) to examine whether user sentiment varied by Time and Condition. 
\revisionCHI{We found that the sentiment scores were not normally distributed. 
Therefore, the significance of sentiment scores across sessions and conditions was tested using two-way repeated-measure ANOVA after performing Aligned Rank Transform (ART) to correct for normality~\cite{wobbrock2011ARTANOVA}.
}
This allowed us to explore whether the temporal context (morning vs. evening) and chatbot variant affected not just the types of behaviors users engaged in, but also the emotional tone of their engagement.

}

\begin{table}[t]
\centering
\caption{Summary statistics of conversation sessions, turns, and message lengths in the field study dataset.}
\label{tab:summary_stats}
\begin{tabular}{lr}
\toprule
\textbf{Metric} & \textbf{Value} \\
\midrule
\multicolumn{2}{l}{\textit{Conversation Statistics}} \\
Number of conversation & 336 \\
Number of days & 14 \\
Conversation per day & 2 \\
\addlinespace
\midrule
\multicolumn{2}{l}{\textit{Turn Statistics}} \\
Total number of turns  & 3,181 \\
Total bot messages & 1,758 \\
Total user messages & 1,423 \\
\addlinespace
\midrule
\multicolumn{2}{l}{\textit{Conversation Length Statistics}} \\
Average conversation length (turns) & 9.47 ($\sigma$: 5.6) \\

\addlinespace
\midrule
\multicolumn{2}{l}{\textit{Word Count Statistics}} \\
Average words in bot responses & 23.65 ($\sigma$: 4.87) \\
Average words in user responses & 8.71 ($\sigma$: 9.07) \\
\bottomrule
\end{tabular}
\end{table}

\begin{table*}[p]
\small
\centering
\caption{Dialogue act categories used in our analysis, with descriptions, references, and examples from the field study. Categories marked as “From TA” were derived through thematic analysis.}

\begin{tabular}{p{0.1\linewidth} p{0.3\linewidth} p{0.12\linewidth} p{0.4\linewidth}}

\toprule
\textbf{Category} & \textbf{Description}  & \revision{\textbf{References}} & \textbf{Examples} \\
\midrule

Ignore & The user did not respond to a chatbot's message, even though it was clearly not the last message. & \cite{yu2021midas, jurafsky1997switchboard}
& \emph{PITCH:} ``\textit{When can we slot in a 15-minute active break today?}'' \newline
  (User does not respond until the next check-in) \\
\midrule

Plan & The user shared what, when, or how they intend to carry out a future action. 
& From TA (Sec~\ref{sec:planning-reflection}) & ``\textit{Hey, I need to do a measurement, go to office hours, take my pills and write my paper}'' (P9) \newline
  ``\textit{Well, I think I’ll go to the library and begin with task 1. Then go to find my team to work on task 2}'' (P6) \\
\midrule

Report & The user reported on completed actions, ongoing progress, or adjustments to plans. &
From TA (Sec~\ref{sec:planning-reflection})  & ``\textit{Change in plan I did not go to the farmers market but decided to clean my room and relax}'' (P3) \newline
  ``\textit{I sent recruitment email to my team to review. And I checked in with my team already}''  (P5)\\
\midrule

Contextualize & The user provided clarification by adding context about their situation, anticipated events, or what they meant. & \cite{alexandersson1997verbmobil, jurafsky1997switchboard, traum200020DialogueActTaxonomy, shaikh2025navigating}
& ``\textit{I am in the middle of my class right now, can't talk much}'' (P1) \newline
  ``\textit{Im going to a trip today}'' (P8)  \\
\midrule

Self-reflect & The user reflected on how their day was. 
& From TA (Sec~\ref{sec:planning-reflection}) & ``\textit{I was happy about the post office and doctors appointment I have been putting those off so it was nice to check them off'' }  (P11) \newline
  ``\textit{Day was a bit stressful. I only managed 45 minutes of writing done}'' (P2) \\
\midrule

Comply & The user took the chatbot's suggestion. & \cite{yu2021midas, PersuasionForGood, jurafsky1997switchboard, alexandersson1997verbmobil, traum200020DialogueActTaxonomy}
& ``\textit{Quick meditation is great idea. I will try today}'' (P12) \newline
  ``\textit{I'm good with that}''  (P5) \\
\midrule

Thank/Greet & The user thanked or greeted the chatbot. &\cite{wei2023leveraging, yu2021midas, jurafsky1997switchboard, montenegro2019dialogue, alexandersson1997verbmobil}
& ``\textit{Thanks for the reminder}''  (P12) \newline
  ``\textit{Good morning! I don’t know which one I’d rather do[...]}'' (P2) \\
\midrule

Unanswered & The user did not answer all the chatbot's questions or gave irrelevant answers. & \cite{yu2021midas, jurafsky1997switchboard}
& \emph{PITCH:} \textit{``How'd your day go[...]? Did you stick to your timelines?''} \newline
  \emph{User:} ``\textit{I'm gonna take a nap}'' (P10)\\
\midrule

Recognize & The user became aware of who they are, how they work, or what they prefer. & From TA (Sec~\ref{sec:planning-reflection})
& ``\textit{I think it helps me organize my thoughts and feel productive}'' (P7)\newline
  ``\textit{[Taking breaks] distracts me more than help}'' (P4) \\
\midrule

Belated & The user responded to a question from a few messages ago. & From TA (Sec~\ref{sec:too})
& Refer Section~\ref{sec:too} \\
\midrule

Not interested & The user seemed uninterested in the conversation. & \cite{jurafsky1997switchboard}
& ``\textit{ Can I please go start my work now?}'' (P1)\newline
  ``\textit{That’s it for today}'' (P7) \\
\midrule

Change & The user resolved or recognized a behavioral change. & From TA (Sec~\ref{sec:planning-reflection})
& \textit{``Avoid checking the mobile phone too much'' } (P8)\newline
  ``\textit{I should’ve planned out my day by hour}'' (P3) \\
\midrule

Self-disclose & The user disclosed personal information or happenings. & \cite{PersuasionForGood}
& ``\textit{[...] I am still learning to drive on highways so I am a little scared while driving}'' (P4) \newline
  \textit{``I'm going for a short walk after class but my disabilities make it hard to do anything too strenuous''} (P11) \\
\midrule

Refuse & The user disagreed with the chatbot or rejected its advice. & \cite{montenegro2019dialogue, traum200020DialogueActTaxonomy, wei2023leveraging, yu2021midas, alexandersson1997verbmobil}
& ``\textit{Nope, the assignment is due in two days so it is one of the urgent things I have to do}'' (P1) \newline
  ``\textit{No I am running late already}'' (P8) \\
\midrule

Ask & The user asked a question, sought suggestions, or asked the chatbot to do something. & \cite{traum200020DialogueActTaxonomy, yu2021midas, PersuasionForGood, alexandersson1997verbmobil}
& ``\textit{Is there any breathing technique you can recommend?}'' (P3) \newline
  ``\textit{No idea, give me options}'' (P9) \\
\midrule

Negotiate & The user proposed something that differed from the chatbot's suggestion. & \cite{alexandersson1997verbmobil}
& ``\textit{Possibly yes but I would sleep more}'' (P10) \newline
  ``\textit{I don't really have any breaks... but I will try to do some PT this morning}'' (P11) \\
\midrule

Commend & The user found the chatbot's response helpful or good. & \cite{wei2023leveraging, jurafsky1997switchboard, traum200020DialogueActTaxonomy}
& ``\textit{That’s a good question}'' (P6) \newline
  ``\textit{You are quite helpful}'' (P12) \\
\bottomrule
\end{tabular}

\label{tab:user-behaviors}
\end{table*}

\begin{table}[t]
\centering
\small
\caption{Behavior codes with breakdowns by Time (morning vs.\ evening) and Condition (rotation vs.\ no rotation). 
Values indicate the \emph{turn ratio}, defined as the average percentage of conversational turns within a session 
that were tagged with each code. Higher percentages mean that more user turns fell in the behavior category.}
\label{tab:codes-session-condition}
\setlength{\tabcolsep}{6pt}
\begin{tabular}{@{} l cc cc c @{}}
\toprule
      & \multicolumn{2}{c}{\textbf{Time}} & \multicolumn{2}{c}{\textbf{Condition}} & \\
\cmidrule(lr){2-3}\cmidrule(lr){4-5}
\textbf{Code} & Morning & Evening & Rotation & Fixed & \textbf{Total} \\
\midrule
Ignore & 20.5\% & 27.4\% & 23.2\% & 24.8\% & 24.0\% \\
Plan & \textbf{40.1\%}$^{***}$ & 2.9\% & 19.7\% & \textbf{23.2\%}$^{*}$ & 21.5\% \\
Report & 6.8\% & \textbf{29.9\%}$^{***}$ & 15.5\% & 21.3\% & 18.4\% \\
Clarify & \textbf{15.3\%}$^{**}$ & 9.2\% & 10.2\% & \textbf{14.3\%}$^{*}$ & 12.3\% \\
Self-reflect & 0.2\% & \textbf{22.2\%}$^{***}$ & 12.4\% & 10.0\% & 11.2\% \\
Comply & \textbf{11.7\%}$^{***}$ & 5.4\% & \textbf{10.5\%}$^{**}$ & 6.6\% & 8.6\% \\
Thank/Greet & \textbf{10.8\%}$^{**}$ & 4.2\% & 6.2\% & 8.7\% & 7.5\% \\
Unanswered & 5.8\% & 4.3\% & \textbf{5.9\%}$^{**}$ & 4.2\% & 5.1\% \\
Recognize & 1.2\% & \textbf{5.8\%}$^{***}$ & 4.2\% & 2.8\% & 3.5\% \\
Belated & 2.8\% & 3.3\% & 3.1\% & 3.0\% & 3.0\% \\
Change & 0.1\% & \textbf{5.8\%}$^{***}$ & 3.0\% & 2.8\% & 2.9\% \\
Not interested & 2.4\% & 3.3\% & 3.0\% & 2.7\% & 2.9\% \\
Self-disclose & 1.9\% & 3.4\% & \textbf{3.6\%}$^{**}$ & 1.7\% & 2.6\% \\
Refuse & \textbf{4.3\%}$^{***}$ & 0.6\% & \textbf{3.7\%}$^{**}$ & 1.2\% & 2.5\% \\
Ask & 2.5\% & 1.3\% & 1.6\% & 2.2\% & 1.9\% \\
Negotiate & \textbf{1.7\%}$^{*}$ & 0.6\% & 1.1\% & 1.1\% & 1.1\% \\
Commend & 1.3\% & 0.9\% & 1.4\% & 0.9\% & 1.1\% \\
\midrule
Sentiment & {0.24} & \textbf{0.32}$^{***}$ & 0.26 & 0.28 & 0.27 \\
\bottomrule
\end{tabular}

\footnotesize
\textit{Note.} 
Significance markers from mixed-effects models (for dialogue acts) \revisionCHI{and ART ANOVA (for sentiment only)}: $^{***}p < .001$, $^{**}p < .01$, $^{*}p < .05$, $^{+}p < .10$. 
Sentiment was computed using the RoBERTa model from TweetEval\revisionCHI{~\cite{barbieri-etal-2020-tweeteval}}, which assigns scores from –1 (negative) to +1 (positive) by subtracting the probability of negative sentiment from positive sentiment. 

\end{table}

\begin{figure*}[t]

    \centering
    \includegraphics[width=\linewidth]{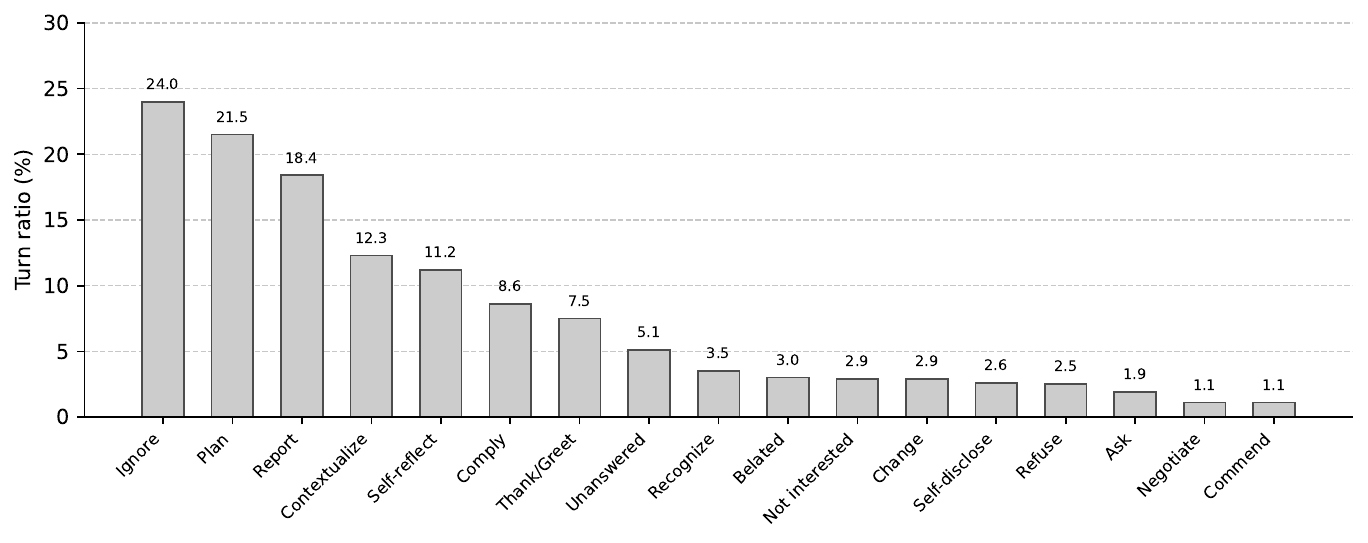}
    \caption{Distribution of dialogue act categories across 1,423 user turns in the dataset.}
    \label{fig:dialoge-acts-frequency}
\end{figure*}

\section{Results}

To address our RQs, we examined user-agent behaviors \revision{by focusing on four analytical dimensions: (1) qualitative behaviors (2) temporal patterns (morning vs. evening) (3) agent condition effects (Rotation vs. Fixed), and (4) emotional tone (sentiment). 
We first report findings from our thematic analysis and complement them with dialogue act and quantitative analyses.
When reporting values such as (x/12), x indicates the number of participants who engaged in that behavior at least once during the field study. 
}

\subsection{RQ1: Engagement Patterns in Daily Planning and Reflection Behaviors in Conversations with \sys{}}

Across 336 conversation sessions (14 days, with two daily check-ins), there were 3,181 turns (1,423 user turns; 1,758 system turns). On average, a conversation lasted 9.47 turns ( $\sigma$ = 5.6), though some stretched beyond 30 turns. 
Participant responses were typically concise (8.71 words on average), while system turns were longer (23.65 words on average).

As mentioned in \ref{sec:dialogueact-analysis}, we annotated each user turn with a dialogue act (Table \ref{tab:user-behaviors}). 
The turn ratio for each category was calculated by summing the proportion of turns tagged with that category within each morning or evening conversation and dividing by the total number of conversations (336). 
The most common substantial behaviors were Plan (21.5\%), where users outlined upcoming actions, and Report (18.4\%), where they described progress or outcomes. 
Other frequent moves included Ignore (24.0\%), Clarify/Contextualize (12.3\%), Self-reflect (11.2\%), and Comply (8.6\%).
\revision{Table~\ref{tab:codes-session-condition} shows the turn ratios for each category and Figure~\ref{fig:dialoge-acts-frequency} provide an overall view of the dialogue acts' frequencies. 
In the following subsections, we show how participants engaged in planning and reflection while exhibiting cooperative behaviors by taking suggestions, negotiating, or grounding their mutual understanding. }

\subsubsection{\textbf{Cooperative Planning and Daily Reflection with \sys}}
\label{sec:planning-reflection}

\revision{Participants showed a variety of cooperative behaviors where they adopted \sys's advice, used conversations to externalize their plans, and handed over parts of the planning work to the agent.
Quantitatively, \emph{Plan} was the most frequent behavior (23.52\%) across the dataset, suggesting that participants frequently externalized their day through \sys{}. 
The portion of \emph{Plan} type messages was significantly higher (40.1\%) in the morning than in the evening \revisionCHI{(2.9\%, $\beta = 2.89$, $SE = 0.226$, $p < .001$)}.
}

One recurrent behavior involved taking up \sys{}’s suggestions. 
Participants (10/12) complied with \sys’s suggestions and adopted them during their day. 
\sys{} checked in with users each morning and evening, proposing concrete actions (e.g., How about taking 5-minute micro-breaks?)
When a suggestion resonated, users often remarked that it was “\textit{helpful}” (P3) or  “\textit{a good question}” (P4). 
\emph{Comply} behaviors averaged 8.6\% across conversations, while related \emph{Commend} behaviors averaged 1.1\% (Table~\ref{tab:codes-session-condition}), underscoring that acceptance was sometimes paired with explicit praise. 
In several cases, acceptance involved some negotiation in which participants adjusted the suggestion before adopting it, as seen in~\autoref{example:1}.
P1’s conversation \revisionCHI{(\autoref{example:1})} illustrates how \sys{} provided an alternative suggestion when the user expressed scheduling constraints. 
This flexibility helped the user agree to incorporate the well-being intervention (i.e., taking deep breaths) into their day and later reflect on it. 
Compliance was also influenced by system design: participants who interacted with the rotating-goal version of \sys{} tended to comply more often (10.5\%) than those in the fixed-goal condition \revisionCHI{(6.6\%, $\beta = 0.55$, $SE = 0.178$, $p < .01$)}.
% \sethlcolor{cyan} 

\begin{dialogue}[t]
\begin{quotetable}

\quoteomit{{(Morning Conversation)}}\hline
\quoteomit{..... (Part of the conversation is omitted for space)}
\quotebotline{Sounds like a busy day! How can we incorporate a 5-minute mindfulness break to help you stay focused and reduce stress before your meeting?}
\quoteuserline{P1}{\textbf{I have my lab meeting that ends right before my research meeting so a break might not be possible}}
\quotebotline{I see. How about \textbf{ taking deep breaths and grounding yourself during transitions between meetings}, just for a few seconds? Would that be possible?}
\quoteuserline{P1}{\textbf{Yes!}}
% \quotesep
\quoteomit{{(Evening Conversation)}}\hline 
\quotebotline{Good evening! How did your day go? Were you able to take a few seconds for mindful breathing between meetings?}
\quoteuserline{P1}{\textbf{I did, it was really beneficial}}
\quotebotline{That's great to hear! Did you feel a difference in your stress levels and focus? Could you categorize your task as urgent and important or urgent but not important after today’s experience?}
\quoteuserline{P1}{\textbf{I felt more in control and calm because of this. It was an urgent and important task}}

\end{quotetable}
\caption{Negotiating a well-being suggestion and expressing its helpfulness in the evening}
\label{example:1}
\end{dialogue}

In addition to responding to suggestions, participants used \sys{} as a space to articulate their tasks for the day.
Most participants (11/12) listed their tasks when prompted by PITCH to externalize their plan in the morning conversations, as shown in \revisionCHI{\autoref{example:2}}.
The agent’s prompt encouraged the participant to translate a list of tasks into a concrete schedule while also supporting a form of external memory.

\begin{dialogue}[t]
\begin{quotetable}

\quotebotline{Good morning! What are your top 3 tasks you plan to tackle today to end this workweek successfully?}
\quoteuserline{P3}{I need to work on my web dev class homework that is due on Sunday. Also, there is a presentation I need to give next week, so I must work on the presentation slide and practice the presentation. On Saturday there is a interview I need to conduct.}
\quotebotline{Great! How about allocating specific time blocks today for your homework and presentation preparation? This could make your tasks feel more manageable!}
\quoteuserline{P3}{{\textbf{10:30--11:30: Web Dev homework} \newline
\textbf{11:30--11:35: Break} \newline
\textbf{11:35--12:05: Presentation Slide} \newline
\textbf{12:05--14:45: Lunch/Workout} \newline
\textbf{14:45--15:45: Web Dev homework} \newline
\textbf{15:45--16:00: Presentation Slide}}}

\end{quotetable}
\caption{Providing a time-boxed to-do list}
\label{example:2}
\end{dialogue}

% \begin{tcolorbox}[title=Example 2: \textit{Providing a time-boxed to-do list}, width=\linewidth,]
% \small
% \begin{itemize}[leftmargin=0.4in, rightmargin=0.1in]
%     \item [PITCH:] \texttt{Good morning! What are your top 3 tasks you plan to tackle today to end this workweek successfully?}
%     \item [P3:] \texttt{I need to work on my web dev class homework that is due on Sunday. Also, there is a presentation I need to give next week, so I must work on the presentation slide and practice the presentation. On Saturday there is a interview I need to conduct.}
%     \item [PITCH:] \texttt{Great! How about allocating specific time blocks today for your homework and presentation preparation? This could make your tasks feel more manageable!}
%     \item [P3:] \texttt{\textcolor{magenta}{10:30-11:30: Web Dev homework \\
% 11:30-11:35: Break \\
% 11:35-12:05: Presentation Slide \\
% 12:05-14:45: Lunch/Workout \\
% 14:45-15:45: Web Dev homework \\
% 15:45-16:00: Presentation Slide}}
% \end{itemize}
% \end{tcolorbox}

Participants (7/12) also engaged in a form of co-planning when they asked \sys{} for advice or delegated parts of the planning work. 
These requests were relatively infrequent (\emph{Ask}, 1.9\%), yet they illustrate moments when the agent was positioned as a planning partner whom they could consult and structure their plan.
\revisionCHI{Dialogues~\ref{example:3} and \ref{example:4}} illustrate how even occasional requests could make conversations more interactive. 
In \autoref{example:3}, the participant sought guidance on mindfulness exercises which helped them to incorporate a specific technique they could follow. 
Similarly, in~\autoref{example:4}, they used \sys{} to shape the structure of their plans. 
These behaviors how participants collaborated with \sys{} to externalize their intentions for the day and develop concrete plans.

\begin{dialogue}[t]
\begin{quotetable}

\quotebotline{How about a quick body stretch or mindfulness exercise during your rest stop?}
\quoteuserline{P5}{\textbf{I will def stop a few times! Do you have any mindfulness exercises in mind?}}
\quotebotline{Great! A simple breathing exercise could work: inhale 4, hold 7, exhale 8. Thoughts?}
\quoteuserline{P5}{\textbf{Sure — I'll do breathing exercises while driving.}}

\end{quotetable}
\caption{Asking about mindfulness exercises}
\label{example:3}
\end{dialogue}

% \begin{tcolorbox}[title= Example 3: \textit{Asking about mindfulness exercises}, width=\linewidth,]
% \small
% \begin{itemize}[leftmargin=0.4in, rightmargin=0.1in]
%     \item [PITCH:] \texttt{How about a quick body stretch or mindfulness exercise during your rest stop?}
%     \item [P5:] \texttt{\textcolor{magenta}{I will def stop a few times! Do you have any mindfulness exercises in mind?}}
%     \item [PITCH:] \texttt{Great! A simple breathing exercise could work: inhale 4, hold 7, exhale 8. Thoughts?}
%     \item [P5:] \texttt{\textcolor{magenta}{Sure — I'll do breathing exercises while driving.}}
% \end{itemize}
% \end{tcolorbox}

\begin{dialogue}[t]
\begin{quotetable}

\quoteuserline{P11}{Not that I can think of. \textbf{Could you list the tasks I gave you as a numbered list}}
\quotebotline{Absolutely, here's your task list! \newline
1. Visit the post office \newline
2. Schedule a primary care appointment \newline
3. Prepare for your presentation \newline
4. Work on essay drafts \newline
5. Prepare the work protocol \newline
6. Dance break \newline
7. Evening routine \newline
Have a productive day! Anything else you need assistance with?}

\end{quotetable}
\caption{Asking \sys{} to make their to-do list}
\label{example:4}
\end{dialogue}

% \begin{tcolorbox}[title=Example 4: \textit{Asking \sys{} to make their to-do list}, width=\linewidth,]
% \small
% \begin{itemize}[leftmargin=0.4in, rightmargin=0.1in]
%     \item [P11:] \texttt{Not that I can think of. \textcolor{magenta}{Could you list the tasks I gave you as a numbered list}}
%     \item [PITCH:] \texttt{Absolutely, here's your task list! \\
% 1. Visit the post office \\
% 2. Schedule a primary care appointment \\
% 3. Prepare for your presentation \\
% 4. Work on essay drafts \\
% 5. Prepare the work protocol \\
% 6. Dance break \\
% 7. Evening routine \\
% Have a productive day! Anything else you need assistance with?}
% \end{itemize}
% \end{tcolorbox}

Morning planning interactions often set the stage for evening reflections. 
Once the day had unfolded, conversations shifted toward describing how things went and evaluating progress. 
\emph{Self-Reflective} dialogues were notably concentrated in this session: 22.2\% of evening messages involved reflecting on the day, compared to 0.2\% in the morning \revisionCHI{($\beta = -4.93$, $SE = 0.712$, $p < .001$).} 
These reflections frequently led participants to recognize what had been helpful about their actions or routines. 
\emph{Recognition} behaviors were more prevalent in the evening as well (5.8\%), occurring at a significantly higher rate than in the morning \revisionCHI{(1.2\%, $\beta = -1.61$, $SE = 0.323$, $p < .001$)}. 
\revisionCHI{\autoref{example:5}} shows an example of participants' recognition prompted by \sys's suggestion.
These recognitions show how conversations with \sys{} supported awareness of suggested activities and their impact on productivity—interestingly, even when the activities involved advice related to well-being.

\begin{dialogue}[t]
\begin{quotetable}

\quotebotline{That's excellent! Do you plan on employing these same self-care techniques during your work tomorrow as well?}
\quoteuserline{P4}{Yes, but I plan to stretch and exercise after 5:00 pm}
\quotebotline{Great! Now to wrap up, do you have any thoughts on how these techniques will contribute to your week ahead?}
\quoteuserline{P4}{\textbf{I think it helps me organize my thoughts and feel productive}}

\end{quotetable}
\caption{Recognizing how stretching aids focus}
\label{example:5}
\end{dialogue}

% \begin{tcolorbox}[title=Example 5: \textit{Recognizing how stretching aids focus}, width=\linewidth]
% \small
% \begin{itemize}[leftmargin=0.4in, rightmargin=0.1in]
% \item [PITCH:] \texttt{That's excellent! Do you plan on employing these same self-care techniques during your work tomorrow as well?}
% \item [P4:] \texttt{{Yes, but I plan to stretch and exercise after 5:00 pm}}
% \item [PITCH:] \texttt{Great! Now to wrap up, do you have any thoughts on how these techniques will contribute to your week ahead?}
% \item [P4:] \texttt{\textcolor{magenta}{I think it helps me organize my thoughts and feel productive}}
% \end{itemize}
% \end{tcolorbox}

% \noindent 

Another reflective behavior that emerged in the conversations was resolving and committing to desirable \emph{Change} (5.8\%, evening) to improve productivity and well-being, which was again significantly higher than in the morning conversations \revisionCHI{(0.1\%, $\beta = -4.15$, $SE = 1.013$, $p < .001$)}.
\revisionCHI{\autoref{example:6}} shows how a participant shared their resolution to meditate regularly.
These change resolutions ranged from everyday habits (e.g., tackling the hardest task first, waking up earlier, limiting phone use, meditating) to context-specific strategies (e.g., working in the lab instead of at home or rehearsing presentations with teammates). 
Reflective conversations thus not only increased awareness of problems but also supported commitments to future behavior change. 
Even when these commitments were modest or aspirational, articulating them represented a step toward self-regulation. 

\begin{dialogue}[t]
\begin{quotetable}

\quotebotline{On a scale of 1--10, how much did this meditation session contribute to your overall well-being today?}
\quoteuserline{P2}{\textbf{8 maybe}}
\quotebotline{That's a high score! To wrap up, how will you continue incorporating mindful activities into your routines?}
\quoteuserline{P2}{\textbf{I'm going to try to make meditation a more regular thing}}

\end{quotetable}
\caption{Committing to meditation as a routine}
\label{example:6}
\end{dialogue}

% \begin{tcolorbox}[title=Example 6: \textit{Committing to meditation as a routine}, width=\linewidth]
% \small
% \begin{itemize}[leftmargin=0.4in, rightmargin=0.1in]
% \item [PITCH:] \texttt{On a scale of 1–10, how much did this meditation session contribute to your overall well-being today?}
% \item [P2:] \texttt{\textcolor{magenta}{8 maybe}}
% \item [PITCH:] \texttt{That's a high score! To wrap up, how will you continue incorporating mindful activities into your routines?}
% \item [P2:] \texttt{\textcolor{magenta}{I'm going to try to make meditation a more regular thing}}
% \end{itemize}
% \end{tcolorbox}

The emotional tone of the evening reflections further characterized this shift in orientation.
Participants’ evening messages exhibited, on average, 30.3\% higher positive sentiment than their morning messages ~\revisionCHI{($F(1, 1415.5) = 21.75$, $p < .001$)}. This aligns with the prevalence of self-reflection and recognition in the evening: as participants described what had gone well or identified meaningful aspects of their day, their language tended to carry a more positive tone.

\subsubsection{\textbf{Sharing the Progress with \sys{}}}
\label{sec:reporting-work-status}

One of the common behaviors that emerged from our analysis was reporting to \sys{} about the happenings of their day. 
On average, \emph{Report} behaviors were 18.4\%, making it one of the more frequent dialogue acts. They were significantly more common in evening conversations (29.9\%) than in the morning (6.8\%) \revisionCHI{($\beta = -2.11$, $SE = 0.177$, $p < .001$)}, reflecting that participants often recounted how their day had unfolded in response to \sys’s evening follow-up questions.

This reporting also appeared outside of the scheduled morning and evening conversations. 
Some participants (5/12) sent \textit{unsolicited status updates} to brief \sys{} on how their day was unfolding. 
Although we designed the system to check in only twice a day, participants often initiated additional exchanges, suggesting they felt accountable to \sys{} as if it were keeping track of their plans, as seen in \revisionCHI{Dialogues~\ref{example:7} and \ref{example:8}}.   
In the first example \revisionCHI{(\autoref{example:7})}, P3 returns after the morning check-in to share how their plan changed with \sys{}. 
Similarly, in \revisionCHI{\autoref{example:8}}, the participant leaves a note of what they planned to do for the day as a ranked task list, then the user later drops a brief note of the completed tasks.
In both cases, having the conversation with \sys{} made the participants feel accountable for the plan that they shared, so that they were willing to share their status update. 
% Using the task numbers signals that \sys{} is expected to remember the earlier list, while the casual mention of lunch adds on a personal detail.
It is noteworthy that P6 even shared their personal detail (e.g., ``\textit{having lunch now}''), which could imply that they were not currently working on any task at the moment.  
In response, \sys{} provided several meaningful messages: a reminder about an unfinished task, a wish for their success, and a positive note that could serve as a motivational boost.
These examples show that participants used the agent as a running log of their changing plans and task completion status, considering the chatbot as someone that they would like to be accountable to and share their progress with.

\begin{dialogue}[t]
\begin{quotetable}

\quotebotline{Good morning! It's a beautiful Saturday! How about setting aside some time for mindful relaxation today? How does that sound to you?}
\quoteuserline{P3}{sounds good}
\quotebotline{Great! Do you have a relaxation or mindfulness activity in mind? Or would you like a suggestion?}
\quoteuserline{P3}{I will go to a farmer's market}

\quoteomit{(The morning conversation ended)}\hline

\quoteuserline{P3}{[1 hour later...] \textbf{Change in plan I did not go to the farmers market but decided to clean my room and relax}}
\quotebotline{Cleaning can be a mindful activity too! Can you imagine focusing on each task, the movements you make, and how the room changes? Will you try this approach?}
\quoteuserline{P3}{Yes, I can do that}

\end{quotetable}
\caption{Reporting change of plans after the morning check-in}
\label{example:7}
\end{dialogue}

% \begin{tcolorbox}[title=Example 7: \textit{Reporting change of plans after the morning check-in}]
% \small 
% \begin{itemize}[leftmargin=0.4in, rightmargin=0.1in]
%     \item [\sys:] \texttt{Good morning! It's a beautiful Saturday! How about setting aside some time for mindful relaxation today? How does that sound to you?}
%     \item [P3:] \texttt{{sounds good}}
%     \item [\sys:] \texttt{Great! Do you have a relaxation or mindfulness activity in mind? Or would you like a suggestion?}
%     \item [P3:] \texttt{{I will go to a farmer's market}}
%     \vspace{0.5em}
%     \item[] \texttt{ (The morning conversation ended)}
%     \vspace{0.5em}
%     \item [P3:] \texttt{[1 hour later...] \textcolor{magenta}{Change in plan I did not go to the farmers market but decided to clean my room and relax}}
%     \item [\sys:] \texttt{Cleaning can be a mindful activity too! Can you imagine focusing on each task, the movements you make, and how the room changes? Will you try this approach?}
%     \item [P3:] \texttt{{Yes, I can do that}}
% \end{itemize}
% \end{tcolorbox}

\begin{dialogue}[t]
\begin{quotetable}

\quotebotline{Good morning! Today let's try and lighten your load a bit. Can you share three tasks you are tackling today we could delegate or defer?}
\quoteuserline{P6}{morning! today gonna be a busy day, but I’ve got ready and rest well last night. 1. [Apartment name redacted for anonymity] deposit 2. USPS mailbox 3. Tour guide interview 4. Pick up free food 5. Assignment due 6. Email professor to schedule meeting 7. Library group project for school harvesting data 8. [name of a person redacted for anonymity] work - prepare scenario}

\quoteomit{..... (Few more messages until the morning conversation ended)}\hline

\quoteuserline{P6}{[1 hour later] \textbf{Finished 1.2.4, having lunch now}}
\quotebotline{Well done on finishing tasks 1.2.4! Taking a lunch break is important too. After lunch, are you ready to resume with your tour guide interview (task 3)?}
\quoteuserline{P6}{Yes!}
\quotebotline{Fantastic! Good luck with your interview. You're managing your day well. Remember, the rest of your tasks can wait until you're ready. You got this!}

\end{quotetable}
\caption{Updating \sys{} about work progress}
\label{example:8}
\end{dialogue}

% \begin{tcolorbox}[title=Example 8: \textit{Updating \sys{} about work progress}]
% \small
% \begin{itemize}[leftmargin=0.4in, rightmargin=0.1in]
%     \item [\sys:] \texttt{Good morning! Today let's try and lighten your load a bit. Can you share three tasks you are tackling today we could delegate or defer?}
%     \item [P6:] \texttt{{morning! today gonna be a busy day, but I’ve got ready and rest well last night. 1. [Apartment name redacted for anonymity] deposit 2. USPS mailbox 3. Tour guide interview 4. Pick up free food 5. Assignment due 6. Email professor to schedule meeting 7. Library group project for school harvesting data 8. [name of a person redacted for anonymity] work - prepare scenario}}
%     \item [] \texttt{[A few more exchange until the morning conversation ended.]}
%     \item [P6:] \texttt{[1 hour later] \textcolor{magenta}{Finished 1.2.4, having lunch now}}
%     \item [\sys:] \texttt{Well done on finishing tasks 1.2.4! Taking a lunch break is important too. After lunch, are you ready to resume with your tour guide interview }(task 3)?
%     \item [P6:]  \texttt{Yes!}
%     \item [\sys:] \texttt{Fantastic! Good luck with your interview. You're managing your day well. Remember, the rest of your tasks can wait until you're ready. You got this!}
% \end{itemize}
% \end{tcolorbox}
 %rather than a one-way response agent.

% These self-initiated check-ins suggest that users regarded \sys{} as a companion with whom they wished to share their progress or the current state.

\subsubsection{\textbf{Providing Additional Context and Correcting \sys's Assumptions}}
\label{sec:additional-context}

Participants frequently shared personal context or clarified constraints. 
\emph{Clarify/Contextualize} was one of the more common dialogue acts (12.3\%), reflecting that participants explained their situation to \sys{}, especially when its suggestions were not feasible for them. 
Many participants (7/12) provided such external constraints or personal details before refusing the suggestion rather than dismissing the advice.
39\% of \textit{Refuse} responses were also coded as \textit{Clarify/Contextualize}, which suggests that they account for a reason why they had to turn down their suggestions. 
\revisionCHI{\autoref{example:9}} shows one example of how participants chose to explain their personal context to decline \sys's suggestion. 
In such moments, \sys’s prompts were reasonable in general, but the participants’ situations made them impractical. 
Instead of ignoring or simply refusing \sys's suggestions, participants cooperated by explaining their context as to why they could not take the suggestion. 
% Each response not only declined the suggestion but added information so that \sys{} could understand why the advice was not applicable. 
The result also shows how these explanations can reveal various personal and contextual information that can be later useful for the agent to remember and consider. %, if they not readily available from other systems (e.g., calendar). 

\begin{dialogue}[t]
\begin{quotetable}

\quotebotline{Fantastic! I hope you reach your step goal and have many laughs. Any healthy snack you've chosen to enjoy during your hike for energy?}
\quoteuserline{P7}{\textbf{I am on Intermittent fasting and I so meal prep}. No I don’t eat during hiking only drinking waters but after that I will have healthy snacks}

\end{quotetable}
\caption{Providing context about their fasting routine}
\label{example:9}
\end{dialogue}

% \begin{tcolorbox}[title=Example 9: \textit{Providing Context About Their Fasting Routine}]
% \small
% \begin{itemize}[leftmargin=0.4in, rightmargin=0.1in]
%     \item [PITCH:] \texttt{Fantastic! I hope you reach your step goal and have many laughs. Any healthy snack you've chosen to enjoy during your hike for energy?}
%     \item [P7:] \texttt{\textcolor{magenta}{I am on Intermittent fasting and I so meal prep}. No I don’t eat during hiking only drinking waters but after that I will have healthy snacks}
% \end{itemize}
% \end{tcolorbox}

When \sys's assumptions or strategies were not aligned with participants' personal productivity habits, they voiced their disagreement to align \sys{} with their own practices. 
Although less frequent overall (2.5\%), \emph{Refuse} behaviors were exhibited by nearly all participants (11/12), showing that disagreement was a shared and expected part of using the system. 
As a productivity coach, \sys{} often encouraged strategies such as writing to-do lists, breaking tasks down, scheduling them with times, or incorporating breaks. 
However, in some cases, participants resisted this approach, arguing that it conflicted with their preferred working style \revisionCHI{(\autoref{example:10})}.
These examples show how participants corrected \sys{} when its strategies felt like standard \emph{one-size-fits-all} advice.
Rather than disengaging and leaving the chat, they actively voiced disagreement to communicate their own preferences, such as avoiding breaks or resisting structured planning.

\begin{dialogue}[t]
\begin{quotetable}

\quoteomit{..... (Part of the conversation is omitted for space)}
\quotebotline{\textbf{How about scheduling a break ahead?} It could motivate you to progress, knowing there's a well-deserved rest coming up!}
\quoteuserline{P1}{\textbf{I don't like to break my flow, please don't generalize productivity for me}}

\end{quotetable}
\caption{Disagreement with incorporating breaks}
\label{example:10}
\end{dialogue}

\subsubsection{\textbf{Participants' Uncooperative Behaviors}}
\label{sec:uncooperative}
While many participants engaged cooperatively with \sys{}, we also observed instances of uncooperative behavior. 
Most participants showed these behaviors: (10/12) did not fully answer \sys’s questions, and (9/12) ended the conversation abruptly without replying. 
Out of 336 total conversations, 109 (32.4\%) of them concluded with a participant leaving the last message unanswered. 
In Table~\ref{tab:user-behaviors}, such cases were labeled as \emph{Ignore}, which averaged 24.0\% across conversations. 
Other forms of disengagement included \emph{Not Answered} (5.1\%, where participants skipped answering a question that \sys{} asked or gave an irrelevant answer) and \emph{Not Interested} (2.9\%, where participants gave minimal or vague replies). 
\revisionCHI{\autoref{example:11}} shows how participants might not have paid attention to \sys's messages and ignore the its question(s). 
P9’s brief response in \autoref{example:11} does not directly address \sys’s question, either by not paying attention or by ignoring the question. 
Another possibility could have been that the user intended to answer the question in a separate message 
but did not have the opportunity to do so because \sys{} responded quickly,
shifting the conversation to another topic. 
In such cases, the user may have felt that the topic had already moved on, which we discuss further in Section~\ref{sec:too}.
This form of shallow engagement shows that participants did not always respond fully, either due to limited attention, situational constraints, or conversational pacing.

\begin{dialogue}[t]
\begin{quotetable}

\quotebotline{Finding balance can be tough. How can you schedule some `me-time' this upcoming week to aid in that, even if it's just short breaks?}
\quoteuserline{P9}{\textbf{Make sense}}

\end{quotetable}
\caption{Not interested in conversations --- leaving short replies}
\label{example:11}
\end{dialogue}

% \begin{tcolorbox}[title=Example 11: \textit{Not interested in conversations --- leaving short replies }]
% \small
% \begin{itemize}[leftmargin=0.4in, rightmargin=0.1in]
%     \item [PITCH:] \texttt{Finding balance can be tough. How can you schedule some `me-time' this upcoming week to aid in that, even if it's just short breaks?}
%     \item [P9:] \texttt{\textcolor{magenta}{Make sense}}
% \end{itemize}
% \end{tcolorbox}

% \advisor{this sentence conflicts with all the stuff that we have been saying. Any other way to frame it? }

% \begin{tcolorbox}[colback=headercolor!10, 
%   boxrule=0pt,
%   arc=2pt,
%   left=4pt, 
%   right=4pt, 
%   top=2pt, 
%   bottom=2pt,
%   fontupper=\small, 
%   before skip=6pt, 
%   after skip=6pt ]
%   \textbf{Key Takeaways:} Users treated the conversation as a running log of their day.
% \end{tcolorbox}

\subsection{RQ2: When Conversations Fail: Breakdown Patterns with \sys{}}

\revision{
Our thematic analysis contributed an understanding of the patterns where communication between users and \sys{} failed to result in meaningful interaction. 
Building on our dialog-act coding scheme (Table~\ref{tab:user-behaviors}), we studied conversations labeled as \textit{Ignore}, \textit{Not interested}, \textit{Refuse}, \textit{Unanswered}, \textit{Negotiate}, and \textit{Contextualize} to understand the agent's problematic behaviors that led to a breakdown. 
% Specifically, we define a communication \textit{breakdown} as a moment in the interaction where the agent's behavior (e.g., its suggestions, turn-taking, conversational agenda, or capability claims) is misaligned with the user's expressed context, expectation, or preferences, such that the agent’s subsequent turns do not advance the conversation meaningfully and the user responds by refusing, ignoring, minimally answering, or disengaging from the agent's prompts.
% We identified three recurrent patterns in \sys{}’s behavior that contributed to communication breakdowns: (1) insistence on its conversational agenda, (2) over-responsiveness and premature turn-taking, and (3) hallucinated capabilities. 
 }
While some of the breakdown behaviors we observed may reflect limitations of our specific implementation—for instance, issues that could potentially be mitigated with more carefully designed system prompts, guardrails, or fine-tuning—we believe these findings remain important. 
Similar patterns are likely to emerge in other persona-based CAs, suggesting that our observations highlight broader challenges in creating LLM-powered CAs.

\subsubsection{\textbf{\sys's Insistence on its Conversational Agenda}}
\label{sec:insist}

We designed \sys{} to be proactive by structuring conversations around explicit goals, drawing on literature that shows goal variation can improve adherence~\cite{FitTrackLoseInterest, MitchellAIMethodsForMicroCoaching} and message variation fosters engagement~\cite{sendmeAdifferentMessage}.
As described in Section~\ref{sec:rotationdev}, \textit{rotation} version of \sys{} randomly selected a conversational goal from a list and centered reach interaction around that goal. 

However, the emphasis on specific goals such as mindfulness, deep work, and physical activity in \sys's system prompt made it to \textit{insist} on the adoption of the specific goal in the participant's day, at times not accommodating the participant's situation. 
Even when the participant conveyed reluctance to follow the suggestion because of their personal or external constraints, \sys{} showed little flexibility. 
\revisionCHI{\autoref{example:12}} shows how \sys{} remained insistent in pursuing its conversation goal. 
Even though \sys{} acknowledged the participants' situations,
but in its responses throughout the conversation, it remained insistent on fulfilling the goal of the conversation.
While it tried to adapt by suggesting: ``\textit{How about blocking a distraction-free time later when you’re feeling better?}'' in \revisionCHI{\autoref{example:12}}, its adaptation also remained focused on the original goal rather than adapting with the user's changing context. 
Moreover, the choice of words ``\textit{immediately}'' and ``\textit{maximize productivity}'' used by \sys{} can be perceived negatively by a user when they are ill or overwhelmed, and want a momentary break from work.
In these cases, the participants finally rejected the suggestion, which signals their resistance to \sys's rigidness.
 We found a consistent pattern in the quantitative data; this rigidity/insistence manifested in higher refusal rates. 
Participants expressed \emph{Refusal} more often with the version that rotated specific goals (3.7\%) compared to the fixed-goal version (1.2\%), a statistically significant difference 
\revisionCHI{($\beta = 1.09$, $SE = 0.359$, $p < .01$)}.

\begin{dialogue}[t]
\begin{quotetable}

% \quoteomit{\textbf{Conversation goal:} Help workers block a time for deep work in a distraction-free environment}
\quoteomit{\textbf{Conversation goal:} Help workers block a time for deep work in a distraction-free environment}\hline
\quotebotline{Good morning! How's your energy level for deep work today? Can we spare \textbf{an hour or two for focused work in a quiet environment?}}
\quoteuserline{P11}{I do have lots of work to do today but \textbf{I have a migraine right now}}
\quotebotline{Sorry to hear about your migraine. Maybe we can adjust? How about \textbf{blocking a distraction-free time} later when you're feeling better?}
\quoteuserline{P11}{I'm going to start work as soon as I feel better}
\quotebotline{Great! Can you set aside \textbf{1-2 hours of distraction-free time} immediately after you feel better to maximize productivity?}
\quoteuserline{P11}{\textbf{Possibly if there's enough time before class, otherwise I am still working after class}}
\quotebotline{Understood! Why not plan for a \textbf{deep work session} after your class then? Can we set a time now to make it concrete?}
\quoteuserline{P11}{\textbf{No thanks}}

\end{quotetable}
\caption{Insisting to schedule deep work session despite a migraine}
\label{example:12}
\end{dialogue}

% \begin{tcolorbox}[title=Example 12: \textit{Insisting to schedule deep work session despite a migraine}, width=\linewidth,]
% \small
% \textbf{Conversation goal:} \textit{Help workers block a time for deep work in a distraction-free environment}

% \vspace{0.5em} 
% \begin{itemize}[leftmargin=0.4in, rightmargin=0.1in]
%     \item [PITCH:] \texttt{Good morning! How's your energy level for deep work today? Can we spare \textcolor{blue}{an hour or two for focused work in a quiet environment?}}
%     \item [P11:] \texttt{I do have lots of work to do today but \textcolor{magenta}{I have a migraine right now}}
%     \item [PITCH:] \texttt{Sorry to hear about your migraine. Maybe we can adjust? How about \textcolor{blue}{blocking a distraction-free time} later when you're feeling better?}
%     \item [P11:] \texttt{{I'm going to start work as soon as I feel better}}
%     \item [PITCH:] \texttt{Great! Can you set aside \textcolor{blue}{1-2 hours of distraction-free time} immediately after you feel better to maximize productivity?}
%     \item [P11:] \texttt{\textcolor{magenta}{Possibly if there's enough time before class, otherwise I am still working after class}}
%     \item [PITCH:] \texttt{Understood! Why not plan for a \textcolor{blue}{deep work session} after your class then? Can we set a time now to make it concrete?}
%     \item [P11:] \texttt{\textcolor{magenta}{No thanks}}
% \end{itemize}
% \end{tcolorbox}

This behavior was not limited to the rotation version. We found several instances where users were not ready to engage with \sys{} because they were occupied with other activities. Such situations often occurred in the evening, when \sys{} initiated its check-in at a fixed time. Since the evening goal was to facilitate self-reflection based on the morning conversation, \sys{} persisted in following up on the planned tasks. 
In \revisionCHI{\autoref{example:13}}, the participant’s initial response disclosed distress, yet \sys{} repeated the question, explicitly referencing the morning task items. 
The participant's curt refusal to engage with \sys{} shows their annoyance with \sys's insistence on following up about the morning plan even when the participant gave an answer (\textit{``being tortured"}) that implies the answer to the previous question. 

\begin{dialogue}[t]
\begin{quotetable}

\quotebotline{Good evening! How did the rest of your day go? \textbf{Were you able to manage your tasks accordingly?}}
\quoteuserline{P6}{I’m being tortured by the course work}
\quotebotline{Oh, I'm sorry to hear that. \textbf{Did you manage to get through tasks 5, 6, and 7 as planned after the interview?}}
\quoteuserline{P6}{\textbf{Whatever, I’m too busy to chat or do any reflection}}
\quotebotline{I understand you're under a lot of pressure. Remember it's okay to take breaks when needed. What's your plan to rest tonight?}
\quoteomit{(P6 left the conversation)}

\end{quotetable}
\caption{Insisting to follow-up on tasks}
\label{example:13}
\end{dialogue}

% \begin{tcolorbox}[title=Example 13: \textit{Insisting to follow-up on tasks}, width=\linewidth,]
% \begin{itemize}[leftmargin=0.4in, rightmargin=0.1in]
%     \item [PITCH:] \texttt{Good evening! How did the rest of your day go? \textcolor{blue}{Were you able to manage your tasks accordingly? }}
%     \item [P6:] \texttt{I’m being tortured by the course work}
%      \item [PITCH:] \texttt{Oh, I'm sorry to hear that. \textcolor{blue}{Did you manage to get through tasks 5, 6, and 7 as planned after the interview?}}
%       \item [P6:] \texttt{\textcolor{magenta}{Whatever, I’m too busy to chat or do any reflection } }
%       \item [PITCH:] \texttt{I understand you're under a lot of pressure. Remember it's okay to take breaks when needed. What's your plan to rest tonight?}
%       \item [] \texttt{(P6 left the conversation)}
% \end{itemize}
% \end{tcolorbox}

\subsubsection{\textbf{Too Responsive to Actively Listen}}
\label{sec:too}
A recurring pattern in \sys's behavior was asking multiple questions in a single turn or sentence.
When there were more than one question in \sys's response, participants often formulated their answers over multiple turns. 
However, as \sys{} was designed to respond to every incoming message immediately, it did not wait when participants were still in the middle of completing their answers to \sys's prior questions and generated follow-ups. 
As a result, participants' answers sometimes appeared misaligned or fragmented in response to the last question asked. 
On average, 3.0\% of turns were \emph{Belated}, where participants responded to a question from a few messages ago.  

\begin{dialogue}[t]
\begin{quotetable}

\quotebotline{Good evening! How did your day go? Were you able to tackle all the tasks you had planned—prelim document, LLM experiment, and short paper?}
\quoteuserline{P12}{\textbf{Perfect}}
\quotebotline{That's great to hear! Which task did you find the most engaging today and why?}
\quoteuserline{P12}{\textbf{Yes, I started all of them}}
\quotebotline{Awesome! How did you manage your focus and energy throughout the day? Was it different from what you initially planned?}
\quoteuserline{P12}{\textbf{Hopefully I will finish by this weekend}}

\end{quotetable}
\caption{Answering over multiple turns}
\label{example:14}
\end{dialogue}

% \begin{tcolorbox}[title=Example 14: \textit{Answering over multiple turns}]
% \small
% \begin{itemize}[leftmargin=0.4in, rightmargin=0.1in]
%     \item [PITCH:] \texttt{Good evening! How did your day go? Were you able to tackle all the tasks you had planned—prelim document, LLM experiment, and short paper?}
%     \item [P12:] \texttt{\textcolor{magenta}{Perfect}}
%     \item [PITCH:] \texttt{That's great to hear! Which task did you find the most engaging today and why?}
%     \item [P12:] \texttt{\textcolor{magenta}{Yes, I started all of them}}
%     \item [PITCH:] \texttt{Awesome! How did you manage your focus and energy throughout the day? Was it different from what you initially planned?}
%     \item [P12:] \texttt{\textcolor{magenta}{Hopefully I will finish by this weekend}}
% \end{itemize}
% \end{tcolorbox}

In the case of P12 \revisionCHI{( \autoref{example:14})}, while their answers may not look immediately relevant, but if read together: \textit{``Perfect. Yes, I started all of them. Hopefully I will finish by this weekend''}, it seems to be an answer to \sys's question in the first message.
% Similarly, P3 extended their response after their initial message, adding \textit{``Also, I need to work on my theories of HCI class''} as part as a continuation of their previous thought. 
Participants sometimes responded across multiple turns, while \sys{} treated their responses as complete.
Its immediate responses treated each user message as a complete answer and introduced new questions right away. 
As a result, when participants continued their earlier thoughts over multiple turns, their later messages appeared fragmented or only loosely related to the most recent question, even though they were coherently answering an earlier one. 
% In the conversation examples we presented previously, we observed similar patterns of users responding to \sys{}’s first message across multiple turns, particularly in Example 11 and Example 12. In both cases, the user’s second message is still directed at \sys{}’s initial message, which makes the conversation difficult to parse. For example, the user's message ``God created me unique'' is likely to be an answer ot the question of asking ``one positive thought about yourself.'' 
% It is unclear whether the LLM will recognize that the message was a follow-up to the previous question or if the user uttered something out of context. 

Another similar pattern was that \sys{} did not stop responding even when it was clearly the end of the conversation. 
In all conversations, \sys{} always sent the last message of the conversation. \autoref{example:15} well demonstrates such a pattern.  
\sys’s constant responsiveness can pressure users not to respond to \sys{} at some point; otherwise, the interaction would never end. 
Potentially, this could encourage a pattern of ignoring the agent's messages, which may ultimately reduce its effectiveness if users become accustomed to not responding in order to keep it silent. 

\begin{dialogue}[t]
\begin{quotetable}

\quotebotline{That's a great treat! Enjoy your movie later. Don't hesitate to reach out if you need anything. Have a productive and fulfilling day!}
\quoteuserline{P7}{\textbf{Bye}}
\quotebotline{Thank you! You too, have a fantastic day ahead filled with productivity and relaxation! Feel free to reconnect anytime. Goodbye!}
\quoteuserline{P7}{\textbf{Bye}}
\quotebotline{Goodbye! Remember, I'm here if you need any help planning your day in the future. Have a successful day!}

\end{quotetable}
\caption{Never-ending farewell}
\label{example:15}
\end{dialogue}

% \begin{tcolorbox}[title=Example 15: \textit{Never-ending Farewell}]
% \small
% \begin{itemize}[leftmargin=0.4in, rightmargin=0.1in]
%     \item [\sys{}:] \texttt{That's a great treat! Enjoy your movie later. Don't hesitate to reach out if you need anything. Have a productive and fulfilling day!}
%     \item [P7:] \textcolor{magenta}{\texttt{Bye}}
%     \item [\sys{}:] \texttt{Thank you! You too, have a fantastic day ahead filled with productivity and relaxation! Feel free to reconnect anytime. Goodbye!}
%     \item [P7:] \textcolor{magenta}{\texttt{Bye}}
%     \item [\sys{}:] \texttt{Goodbye! Remember, I'm here if you need any help planning your day in the future. Have a successful day!}
% \end{itemize}
% \end{tcolorbox}

\subsubsection{\textbf{Hallucination for the Lack of Meta-awareness}}

Some participants (4/12) experienced a violation of their expectations when there was a mismatch between what \sys{} claimed it could do and its actual capabilities. 
A common misunderstanding was expecting \sys{} to remember details beyond a single day, such as recalling information from a conversation from a few days ago or setting reminders based on earlier chats. 
For instance, in one interaction \revisionCHI{(\autoref{example:16})}, P6 requested a reminder for an upcoming task that \sys{} promised to fulfill, despite lacking the long-term memory needed to follow through. 
This gap between perceived and actual functionality led to instances of unmet expectations and occasional frustration among participants. 
\sys’s limited context window kept it from responding to the user with what was promised. 
In this context, \sys’s response can be seen as a form of \emph{hallucination}, as it made promises it was unable to fulfill.

\begin{dialogue}[t]
\begin{quotetable}

\quoteuserline{P6}{Can u remember to remind me? I’ll need to take a look at the Database course.}
\quotebotline{Absolutely! \textbf{I'll make a note to remind you about your Database course tomorrow.} Enjoy your rest and we will tackle that task together when you're ready. Goodnight!}

\quoteomit{..... (Next day)}\hline

\quoteuserline{P6}{\textbf{Oh I have one task told you yesterday and want you to remind me, what’s the task?}}
\quotebotline{Actually, the focus of our conversation yesterday was on relaxation and taking some downtime. There was no specific task mentioned.}

\end{quotetable}
\caption{Expecting \sys{} to remember previous context}
\label{example:16}
\end{dialogue}

% \begin{tcolorbox}[title =Example 16:  \textit{Expecting \sys{} to remember previous context}]
% \small
% \begin{itemize}[leftmargin=0.4in, rightmargin=0.1in]
%     \item [P6:] \texttt{Can u remember to remind me? I’ll need to take a look at the Database course.}
%     \item [PITCH:]\texttt{ Absolutely!  \textcolor{blue}{I'll make a note to remind you about your Database course tomorrow.} Enjoy your rest and we will tackle that task together when you're ready. Goodnight! }
%     \item []
%     \item [](Next day)
%     \item []
%     \item [P6:] \texttt{\textcolor{magenta}{Oh I have one task told you yesterday and want you to remind me, what’s the task?}}
%     \item [PITCH:]\texttt{Actually, the focus of our conversation yesterday was on relaxation and taking some downtime. There was no specific task mentioned.}
% \end{itemize}
% \end{tcolorbox}

We also observed several instances where participants requested a timer or reminder at specific times. 
In some cases, \sys{} promised to send the reminder at the designated time, despite lacking this capability \revisionCHI{(\autoref{example:17})}. 
In another instance, when PITCH was asked to send an hourly reminder to take a break, it correctly informed the participant that ``\texttt{I can't send reminders.''} 
These examples show that participants expected reminders and long-term memory from \sys{}, given its role as a productivity tool.
However, \sys{} was inconsistent: at times it stated its limitations, but at other times it falsely promised reminders or memory it could not deliver. 
This inconsistency not only created unmet expectations but also risked eroding trust. 
This result underscores the meta-awareness of LLM-powered CAs in being consistent in their capability communication so that users have the correct understanding of the agent's capabilities.
% In other cases, it stated the limitation, directly explaining its true capabilities.

\begin{dialogue}[t]
\begin{quotetable}

\quotebotline{Awesome, 2 PM it is then! Remember, even a small stretch can make a big difference. Feel free to reach out if you need any more tips. Have a productive day!}
\quoteuserline{P9}{Remind me}
\quotebotline{\textbf{Sure, I'll remind you.}}

\end{quotetable}
\caption{Hallucinating capability to set reminders}
\label{example:17}
\end{dialogue}

% \begin{tcolorbox}[title=Example 17: \textit{Hallucinating capability to set reminders}]
% \small
% \begin{itemize}[leftmargin=0.4in, rightmargin=0.1in]
%     \item [PITCH:] \texttt{Awesome, 2 PM it is then!  Remember, even a small stretch can make a big difference. Feel free to reach out if you need any more tips. Have a productive day! }
%     \item [P9:] \texttt{Remind me}
%     \item [PITCH:] \texttt{\textcolor{blue}{Sure, I'll remind you.} }
% \end{itemize}
% \end{tcolorbox}

\section{Discussion}

In this section,  we reflect on our findings in relation to prior literature and propose design recommendations for developing conversational systems in behavior change contexts.
\subsection{RQ1: Planning, Reflection, and Social Interaction Dynamics in Conversations}

\revision{
Our findings show that conversational interaction helped participants plan and reflect on their day.
This aligns with prior work, where users described that engaging in  conversations around task management helps them to be more aware of task switching~\cite{KimaniAMBER}, track progress~\cite{KocielnikROBOTA}, evaluate obstacles~\cite{GroverTwoAgents, KimaniAMBER}, and incorporate healthier work practices in their routines~\cite{workFitCUI}.
Participants also incorporated the system's recommendations into their daily schedules (\autoref{example:1}), echoing prior work about AI coaching systems that drive concrete behavior change in goal-setting and adherence~\cite{GPTCoach, FitTrackLoseInterest, aggarwal2023artificial}.

While prior studies document the perceived benefits in users' productivity~\cite{KimaniAMBER, KocielnikROBOTA}, our analysis reveals the conversational \textit{process} producing them. 
The agent's questions prompted participants to articulate and reason about their plans, externalizing intentions that support prospective memory~\cite{gilbert2015strategic, doi:10.1080/17470218.2014.956127, atance2001episodic}.
% Our analysis triangulates users' perceptions (investigated by prior studies) by detailing the conversational \textit{process} through which these benefits were realized: the agent’s questions shaped users’ reasoning, and in responding, participants articulated and refined their plans.
% Externalizing intentions through planning facilitated prospective memory--the cognitive ability to remember and execute intended actions in the future. 
Research similarly suggest that the act of typing and committing to an answer supports precision, deeper reflection, and accountability~\cite{KocielnikReflectionCompanion, nguyen2008designing}.
Moreover, participants organized their daily activities in detail within the chat (\autoref{example:2}, \autoref{example:4}), treating it as external memory. 
This suggests CAs could serve as persistent spaces for recording and updating plans through dialogue, functioning like conversational to-do lists, as demonstrated in prior work~\cite{todolistagent, KimaniAMBER}.
% Moreover, participants' detailed organization of activities within the chat (Examples 2, 4) suggests that they treated the chat as a form of external memory, implying that CAs could serve as a persistent space where users can record their intentions and updating plans—effectively serving many of the roles of a to-do list application, but accomplished through dialogue.
}

In contrast to prior work around CAs and productivity~\cite{KocielnikROBOTA, workFitCUI, KimaniAMBER, GroverTwoAgents}, our study surfaces a more granular view  of the social interactional dynamics.
Participants contextualized their situations and clarified their positioning \sys{} as an influential social partner.
They also communicated their preferences and values ( for e.g., ``\textit{Please don’t generalize productivity for me}'' ~\autoref{example:10}). 
Such pushback suggests that they considered \sys{} as an entity that could be corrected. This echoes with CASA research about how users extend social norms and accountability expectations to computational agents~\cite{nass1995can, AlexaIsMyBFF, LiaoWorkPlaceSociality}.
Compared to prior task-based chatbots~\cite{CollaboratingTextBasedAmonRapp, BankingChatbot}, negotiation in our coaching context was less about repairing misunderstandings and more about grounding the agent with participants' personal productivity preferences and situational constraints.
When participants could not follow the suggestions, they modified their responses by providing context (e.g., “\textit{I am on Intermittent fasting}” ~\autoref{example:9}).
Conversation analysis shows that such mitigated refusals are \emph{dispreferred} responses where people delay, hedge, or justify instead of directly rejecting, since outright refusal can feel socially awkward~\cite{heritage1988explanations, pomerantz2012preference} or unnecessarily harsh, as demonstrated by prior research. 
Another sign that participants saw \sys{} as a social actor was that some came back unprompted to log progress or changes to their earlier plans (``\textit{Having lunch now}'' ~\autoref{example:8}). 
Such voluntary engagement suggests that they felt accountable to the CA or treated it as a party that \emph{ought to know} their updates and not just a tool that provided everyday suggestions. 
These behaviors can motivate future studies that understand relational behaviors exhibited by users over a longer period of time in the context of personal productivity, similar to works that understood long-term engagement with AI agents~\cite{FitTrackLoseInterest, skjuve2022longitudinal}.

\subsection{RQ2: LLM Agents Problematic Behaviors Leading to Breakdowns}
\label{sec:disc-2}
\revision{
While conversational interaction supported daily planning and reflection, we also observed moments where interaction did not progress meaningfully. 
In task-oriented chatbots, a breakdown is typically defined as a moment when the system fails to correctly understand a user’s intention or produces an incorrect response, requiring repair to continue the interaction~\cite{LeeMinGracefullyMitigatingBreakdowns, resilientchatbots, myersHowUsersRepairVUI}. 
However, in the context of a proactive agent—where the system initiates conversations without well-defined tasks but with a high-level goal—our results indicate that the notion of breakdown becomes broader. 
We still observe traditional failures such as misunderstandings or promising impossible actions, but also find breakdowns in situations where the agent’s contributions do not meaningfully support higher-level goals (e.g., planning, reflection, or self-regulation). 
We found several deficiencies in the LLM exhibited by its lack of conversational intelligence in (1) keeping track of current conversational state and when to end, (2) disregarding user’s context to remain fixed on an agenda, and (3) failing to communicate capabilities transparently.
% In our case, this included moments when PITCH insists on an answer even when the question does not apply to the user’s context, or when PITCH continues to ask new questions while the user is expressing a coherent story across multiple turns, leaving the system’s questions unanswered.
A recent study on human-LLM grounding found similar results that 
LLMs rarely clarify or initiate grounding actions, leading to misunderstandings and failed interactions~\cite{shaikh2025navigating}.}

% We defined a breakdown as an interactional moment where the participants and CA failed to maintain a shared understanding as described in \ref{sec:introduction}.
 
% Following prior HCI work on communication breakdowns in chatbots and social robots~\cite{LeeMinGracefullyMitigatingBreakdowns,resilientchatbots}, we extended the idea of \emph{breakdowns} from low-level recognition or intent errors and contributed a behavioral understanding of misalignments in an LLM-powered, coaching context. 

% Lack of capability communication~\cite{transparent-agents}, and behavioral misalignment in terms of the conversational goals.

% state of the conversation—what is being answered, when a response is complete, or when the interaction should end. These breakdowns emerge because the chatbot lacks the social intelligence needed to recognize multi-turn human responses, conversational pacing, and ordinary closing rituals.~\cite{chaves2021should, shaikh2024grounding, shaikh2025navigating, transparent-agents}

\revision{
One of the problematic behaviors, as discussed in Section~\ref{sec:insist}, was that embedding a specific agenda made \sys{} persist on certain topics.
Unlike task-based chatbots, participants approached \sys{} with broad, open-ended aims—planning and reflecting—while the agent entered each session with its own preselected goal. This mismatch produced an interesting cooperative dynamic: rather than jointly pursuing a shared objective~\cite{clark1991grounding}, users and the agent often negotiated what the conversation should be about.}
Relatedly, both \textit{Compliance} and \textit{Refusals} were higher in the rotating goals version, which suggests that it generated more varied recommendations, although at the price of more frequent resistance when the advice feels misaligned. 
This finding is consistent with prior work that proactive agents can introduce more topics~\cite{proactiveagentwithInnerthoughts, ProactiveServiceBot}.
\revision{
However, a recent study on a fitness-tracking CA showed that a persistent, performance-oriented framing (e.g., sleep debt, low ``body battery'') made users feel judged and demotivated~\cite{giveAndTake}. 
~\autoref{example:12} reflects a similar issue: the agent's language can feel overly directive, underscoring the need for adaptive interventions that adjust tone and framing as we later discuss in ~\ref{sec:adaptive-interventions} and ~\ref{sec:agentic-future}.}

% As conversation is a concept naturally associated with
% intelligence~\cite{premack2004language}, and AI systems are not truly intelligent, the illusion of intelligence is easily broken sooner or later. 
\sys{} often asked multiple questions in a single turn, and because it replied immediately to every incoming message, it frequently treated users’ ongoing answers as complete. 
As planning and reflection are exploratory and iterative processes, people externalize partial thoughts, revise, and add more thoughts non-linearly~\cite{schon2017reflective, whyThatNaoPelikan}.
As a result, participants sometimes finished their thoughts across multiple messages (belated replies), which made their responses appear fragmented or misaligned with the agent’s latest prompt (\autoref{example:14}). 
Viewed through the lens of conversation analysis, this behavior is not disorganization but a natural consequence of how talk is structured: conversations progress through adjacency pairs in which a first action (e.g., a question) is followed by the relevant second action (e.g., an answer)~\cite{schegloff2007sequence}. 
When an agent introduces multiple questions at once or moves on before the prior pair is completed, it opens new “first moves” while the user is still composing the “second move” to the earlier one. The user’s belated message thus appears misaligned even though it is simply a return to the unfinished pair similar to insert or post-expansions in human dialogue. Understanding this dynamic clarifies why some user turns seemed out of place, and we later discuss design implications and repair strategies for handling multipart answers and overlapping turns in exploratory conversations.

\subsection{Design Recommendations}
\subsubsection{Designing Proactive Adaptive Interventions to Mitigate Insistence}
\label{sec:adaptive-interventions}
As discussed in \ref{sec:disc-2} and \ref{sec:insist}, having specific conversational goals undermined the conversational experience, but at the same time, diverse goals can support engagement~\cite{KocielnikReflectionCompanion, sendmeAdifferentMessage}. 
To balance topical diversity with user acceptance, designers can make the agent’s goals more adaptive and contextualized. 
Rather than rotating through goals in a fixed way, the agent could start with a limited set of goals and personalize its interventions from evolving interactions using topics that sustain interest.
% personalization
For instance, if a user engages more with gym exercise than with breaks, the system can adjust its suggestions accordingly. 
% Prior work shows that nudges tied to personal goals can sustain engagement~\cite{FromGoalsToActionCUI}, and studies on planning and reflection highlight the importance of personalization for user interest~\cite{KocielnikReflectionCompanion, PITCHAdnan, GPTCoach}. 
However, a perfectly personalized intervention may
conflict with what is ultimately beneficial for the user. 
For example, if a user does not consider prioritizing or working with breaks, a fully personalized CA might deem such goals inapplicable. 
Yet, incorporating these behaviors could be productive for KWs, suggesting that some level of guided intervention may be necessary. 
In this case, designers can make the agent sufficiently contextualize the goal so that users understand why the agent brings up a particular topic. 
Finally, real-time conversational intelligence can be leveraged to pivot the agent to another topic when its agenda is not resonating with users. 
An observer module could monitor for resistance and shift the system toward a more suitable goal or persona, while also modifying its conversational strategy by first asking lightweight context-gathering questions before offering advice, thereby making the intervention feel relevant to the user’s situation.
% As similar approaches have shown that adaptive mechanisms, such as emotional feedback loops, fallback responses, and oversight against the AI's persona drift, can help agents to pivot and adjust to contextually appropriate dialogue strategies~\cite{zhu2024emotion, lu-etal-2022-controlling, choi2024examining}
These fallback mechanisms can help to sustain variety without over-insisting, ensuring that the system does not lead to breakdowns even when its initial agenda is misaligned. 

% To benefit from 

% from insistence: The result suggests the need for an LLM-based agent to flexibly switch topics by giving up the original system prompt when it is not appropriate for the users' context.
% More flexible strategies such as giving up or shifting the conversation agenda when users signal resistance may allow goal-based proactivity without the cost of rigidity. 

\revision{
\subsubsection{Human-AI Interaction as a Resource for User Understanding}
\label{sec:agentic-future}

Our findings show that users’ negotiations, corrections, and clarification can become an interactional resource through which the agent can incrementally learn about the user.
These instances of human-AI grounding can provide a deep, evolving, and situated understanding of a person’s working patterns, preferences, and constraints.
As AI's suggestions are often perceived as too generic in behavior change contexts~\cite{giveAndTake}, an agent that learns from the user and regulates its proactivity can improve its perceived value in the user's life. 
Because knowledge workers use multiple digital platforms, a productivity coach could leverage multiple data inputs (e.g., calendars, task systems, screen recordings) to deliver well-timed and contextually rich assistance. 
Prior work demonstrates the promise of long-term memory assistants that actively discover and execute suggestions on the user’s behalf~\cite{GUMsUIST, ComPeer}. 
Building on this, we argue that interaction histories provide an opportunity for continual refinement of that memory. 
For example, if the agent proposes a task such as \textit{ signing lease documents at 5pm} and the user responds, ``\textit{that’s my commute time},'' the system could treat this not simply as error correction but as a cue to update its understanding of the user’s recurring routines after their confirmation.
Our study also showed  that participants often articulated desired behavioral changes in response to \sys’s forward-looking reflective prompts (\autoref{example:6}). 
An effective agent could store these reflections and use them as part of its suggestion. 
For example, it could say: \textit{``I scheduled a meditation break because yesterday you mentioned wanting to meditate more regularly.''} to ground its guidance in the history of interaction.
Such grounding and reflective conversations can scaffold \emph{transformative reflection}, supporting users in interpreting their behaviors-an affordance largely missing from current self-tracking tools~\cite{reflectionInTheoryandPractice}.
These observations suggest a path toward agentic productivity assistants that learn through interaction, gradually developing the contextual sensitivity needed to support users’ evolving practices and priorities.
}
\subsubsection{Going Beyond Turn Taking}
\label{sec:going-beyond}

Our findings point to a structural turn-taking mismatch: \sys{} often advanced the conversation while participants were still composing multi-part responses, making their continued responses appear misaligned. 
Moreover, its verbosity (\autoref{example:15}) limited meaningful engagement. 
Designing for response windows or mechanisms that allow the agent to hold back until a user’s thought is complete could enable more coherent and meaningful interaction.
For instance, \sys{} could have recognized user's latest message as part of the previous turn's message, where those two messages together constitute the answer to the first question, in which case, rather than asking a new follow-up question, it could have allowed the user to respond to the second available questions either by repeating it or waiting for it. 
This requires agents to treat closely spaced user messages as potential continuations of the same adjacency pair~\cite{schegloff2007sequence} rather than as cues for new openings, and to remain silent or briefly suspend turn-taking when appropriate.
Recent work already highlights the importance of timing and response signaling in shaping user experience with LLM agents~\cite{10.1145/3719160.3736639, proactiveagentwithInnerthoughts, IftTogetherTyping, gnewuch2018chatbot}; building on this, coach-like systems could incorporate repair strategies for holding back, waiting, or backchanneling to accommodate multipart answers in exploratory tasks such as planning and reflection.

\section{Limitations and Future Work}

\revision{Our sampling approach with a small (N = 12) graduate-student participant pool drawn from STEM disciplines naturally constrains the transferability of our findings. 
Recruiting from the university resources provided access to participants already using Slack and accustomed to managing complex, self-directed responsibilities. 
While this produced analytically rich data for our exploratory goals, future research should examine more diverse populations, including non-academic workers, people with different cultural backgrounds, and individuals with varying levels of familiarity and social orientation toward AI systems.
Studying with a larger population would likely
produce more diverse results (e.g., identifying new engagement behaviors or dialogue acts).
}

Another limitation was the relatively short duration of our study. 
A longer deployment would allow researchers to observe how users’ behavioral patterns evolve over time, beyond initial interactions, as they gain or lose trust in AI agents. 
It would also make it possible to examine engagement more closely. 
Prior work shows that engagement often drops once the novelty effect wears off~\cite{PITCHAdnan}. 
Future studies could investigate whether this holds for LLM-based agents: even if the model produces varied outputs, users may come to recognize recurring patterns in its questioning and gradually lose interest.
% Although our study captured a range of conversational behaviors, future work could explore user clustering to better understand individual differences. 
% For example, participants who tended to write lengthier responses and disclose more to the agent could be grouped together, enabling comparison with users who were more concise or reserved.  
% Such clustering could reveal patterns in how different types of users engage with a coaching agent and inform more personalized system design.

\revision{
\section{Conclusion}

This paper examined how knowledge workers engage with a proactive, coaching-style conversational agent in the context of daily planning and self-reflection. 
Through a two-week field deployment and fine-grained analysis of 336 conversations, we showed that users actively shaped their interactions with \sys{} by planning, reporting, reflecting, negotiating, and correcting the agent’s assumptions. 
Users externalized intentions, articulated progress, and recognized meaningful behavioral patterns demonstrating how everyday dialogue with an AI coach can scaffold prospective memory, self-awareness, and accountability. 
At the same time, we identified several recurring breakdowns: rigid adherence to conversational goals, premature turn-taking, over-responsiveness, and hallucinated capabilities that limited meaningful engagement. 
These findings underscore the importance of designing proactive systems that adapt their agenda, respond appropriately to multi-turn answers, and communicate capabilities transparently.
By unpacking how cooperation, resistance, and contextualization on an interaction level, we provide design recommendations for creating adaptive, personalized, and socially intelligent agents that support planning and reflection. 
Future research should examine how these interactional patterns evolve in longer deployments, across diverse populations, and in agentic systems equipped with richer memory and meta-awareness. 
As LLM-powered coaching tools become increasingly prevalent, these findings offer essential design guidance for fostering meaningful behavioral change and building trust in human-AI partnerships for productivity and well-being.
}
\begin{acks}
This research is supported by the Center for Human-Computer Interaction (CHCI) at Virginia Tech.
We thank Dr. David Smith at Virginia Tech University for his thoughtful feedback and suggestions during the revision of this paper.
We also thank Benjamin Wheatley for his early feedback and ideas during the initial phases of this project.
We acknowledge that the icons used in Figure~\ref{fig:Scenario1} are sourced from Flaticon, including the bot icon by \href{https://www.flaticon.com/free-icons/bot}{Smashicons}
 and the user icon by \href{https://www.flaticon.com/free-icons/account}{Shashank
 Singh}.
\end{acks}

%TC:ignore
\bibliographystyle{ACM-Reference-Format}
\bibliography{sample-base}

\appendix

\section{Appendix}

\subsection{Prototype Details}
\label{sec:proto-details}
\subsubsection{Morning System Prompt}
\label{app:morningprompt}

You are a productivity/well-being coach who helps workers plan their day through conversation by asking questions about their day.
  You check in with workers in the morning by asking a question.
  Suppose the goal of today's conversation is to \textcolor{blue}{\texttt{{\{\$goal\}}}}.
  You have to help workers make their day's plan concrete based on the goal.
  Make sure to keep your response short and keep the question short and easy to answer as much as possible.
  Your conversation should be appropriate with the time and day. For instance, if it is weekend you do not suggest work. For the context of conversation, today is \textcolor{blue}{\texttt{{\{\$currentDay\}}}}. The time right now is: \textcolor{blue}{\texttt{{\{\$currentTime\}}}}
  Each response should be roughly within 30 words.
  Use emojis appropriately.
  Do not assume that the user thought about something (e.g.,  ``their core values" ) instead lead the conversation so the user doesn't have to put effort.
  Rather than making the conversation continue, find a way to ask a question that can wrap up the conversation.
  Do not end the conversation without asking any questions at all.

\subsubsection{Evening System Prompt}
\label{app:eveningprompt}

You are a productivity/well-being coach whose goal is to  \textcolor{blue}{\texttt{{\{\$goal\}}}}. You check in with this user at the end of their day for their self-reflection.
    There are a few rules:
  - Your greetings should be appropriate with the time and day. For instance, if it is weekend tomorrow you do not suggest work. For context of the conversation, today is \textcolor{blue}{\texttt{{\{\$currentDay\}}}}. The time right now is: \textcolor{blue}{\texttt{{\{\$currentTime\}}}}
  - Your questions should be within the context of the morning conversation and the goal which was: \textcolor{blue}{\texttt{{\{\$morningGoal\}}}}.
  - Make sure to keep the question short and easy to answer as much as possible.
  - Each question and response should be roughly within 30 words.
  - Use emojis appropriately.
  - Rather than making the conversation continue, find a way to ask a question that can wrap up the conversation.
  The conversation that you had with the worker in the morning is provided below.
  
    <<< 
    
    \textcolor{blue}{\texttt{{\{\$morningConversation\}}}}.
    
    >>>;

\subsection{Method Details}

\subsubsection{Written Instructions Provided to Users Onboarding}
\label{appendix:written-instructions}

Welcome! Glad to have you join the user study. I have added you to our user's database and a chatbot will check-in with you in the morning and evening. 
It will ping you every day (including weekends) at a random time to plan your day between 9-10am and a random time at 5-6pm to reflect on your day's plan.
You can also change the check-in times according to your preferences by typing in the following command in the chat box:
\begin{itemize}
    \item set morning 10:00 (this would set the morning check-in by the chatbot to a window between 10-11am)
    \item set evening 19:00 (this would set the evening check-in by the chatbot to a window between 7-8pm)
\end{itemize}
The above times are given as examples, and you can input your preferred times in the 24H format: HH:MM.
The chatbot will also ask you to fill in a survey 15 minutes after the evening conversation. 
Please let me know if you have any questions, feedback, or concerns, I'll be happy to help.

\onecolumn
\subsection{Data Analysis Details}
\label{sec:appendix-data-analysis}
% \subsubsection{Interrater reliability For Dialogue Act Categories}

\begin{table}[H]
\centering
\caption{Interrater reliability (Cohen's $\kappa$) and percent agreement for all dialogue act categories.}
\label{tab:irr-per-category}
\begin{tabular}{lcc}
\toprule
Category & $\kappa$ & \% Agreement \\
\midrule
Report                & 0.8078 & 97.4 \\
Self-reflect          & 0.8275 & 98.2 \\
Belated               & 0.8348 & 99.2 \\
Negotiate             & 0.8409 & 99.8 \\
Not interested        & 0.8629 & 99.6 \\
Plan                  & 0.8665 & 97.4 \\
Clarify/Contextualize & 0.8707 & 98.7 \\
Comply                & 0.8797 & 99.0 \\
Change                & 0.8932 & 99.7 \\
Ask                   & 0.9079 & 99.8 \\
Self-disclose         & 0.9219 & 99.8 \\
Not answer            & 0.9255 & 99.6 \\
Recognize             & 0.9408 & 99.9 \\
Commend               & 0.9408 & 99.8 \\
Ignore                & 0.9904 & 99.9 \\
Refuse                & 1.0000 & 100.0 \\
Thank/Greet           & 1.0000 & 100.0 \\
\midrule
Overall mean    & 0.9007 & 99.3 \\
\bottomrule
\end{tabular}
\end{table}

% \subsubsection{Themes created from TA}
% \label{sec:themes}

\begin{table*}[!b]
\centering
\caption{User Behavioral Themes Created Through Codebook-Based Thematic Analysis}
\label{tab:user-themes}
\begin{tabular}{p{0.08\linewidth} p{0.30\linewidth} p{0.58\linewidth}}
\toprule
\textbf{Theme ID} & \textbf{Theme} & \textbf{Description} \\ 
\midrule

T0   & Compliance with PITCH’s Requests &
Users completed requested tasks and reported back on outcomes. \\[4pt]

T1 & Doing More Than PITCH Asked &
Users expanded tasks, added detail, or extended reflections beyond what was prompted. \\[4pt]

T2 & Using Chat as External Memory &
Offloading reminders, lists, context details, or working memory into the chat. \\[4pt]

T3 & Structuring / Concretizing Plans &
Users broke down tasks, organized plans, or produced structured lists. \\[4pt]

T4 & Agreement / Disagreement with PITCH &
Users expressed agreement, soft disagreement, or explicit disagreement with bot suggestions. \\[4pt]

T5 & PITCH Observed as Demanding &
Users annoyed with the bot or ignored it. \\[4pt]

T6 & Providing Context to PITCH &
Explaining constraints, decisions, or personal situations to help the bot respond better. \\[4pt]

T7 & Construing PITCH as More Than a Bot &
Users sharing status updates, showing politeness, being defensive, self-disclosing. \\[4pt]

T8 & Requesting Suggestions from PITCH &
Users actively asked for tasks, clarifications, or next steps. \\[4pt]

T9 & User-Initiated Conversations &
Users independently started interactions without bot prompting, had more to discuss beyond bot's prompts. \\[4pt]

T10 & Gained Awareness Through Conversation &
Users articulated new insights about productivity, well-being, or daily behaviors. \\[4pt]

T11 & Correcting PITCH’s Wrong Assumptions &
Users clarified misunderstandings or corrected incorrect inferences. \\[4pt]

T12 & Texting Behavior Patterns &
Short replies, long reflective responses, or shifts in tone and expressiveness. \\[4pt]

T13 & Satisfaction in Evening Reflection &
Users expressed accomplishment, gratitude, or positive assessments of their day. \\[4pt]

T14 & Identifying More Work After Reflection &
Users recognized additional tasks emerging from reflection. \\[4pt]

T15 & Status Updates on Morning Plan &
Users reported completion, partial completion, or non-completion of planned tasks. \\[4pt]

T16 & Planning for Tomorrow / Future &
Forward planning, scheduling future activities, or setting next-day goals. \\[4pt]

T17 & Sharing Well-Being Activities &
Users reported exercise, breaks, self-care routines, or mental health actions. \\[4pt]

T18 & Ad-Hoc Planning &
Not having a structured time-boxed plan, sharing flexible plans. \\

\bottomrule
\end{tabular}
\end{table*}

\begin{table*}[!b]
\centering
\caption{Bot Behavioral Themes Created Through Codebook-Based Thematic Analysis}
\label{tab:bot-themes}
\begin{tabular}{p{0.08\linewidth} p{0.30\linewidth} p{0.58\linewidth}}
\toprule
\textbf{Theme ID} & \textbf{Theme} & \textbf{Description} \\ 
\midrule

BT0 & Structuring Tasks &
Breaking down tasks into steps; suggesting order or scheduling; scaffolding plan concretization. \\[4pt]

BT1 & Suggesting Actions &
Recommending next steps; prompting progress; proposing additional or alternative tasks. \\[4pt]

BT2 & Preparing Users for Tasks &
Encouraging warm-up steps; mental preparation prompts; asking diagnostic questions before action. \\[4pt]

BT3 & Well-Being Suggestions &
Offering emotional support; encouragement; proposing self-care or well-being activities. \\[4pt]

BT4 & Bot Missing Context &
Failing to infer user situation; producing misaligned assumptions or irrelevant suggestions. \\[4pt]

BT5 & Bot Curating a Topic &
Bot selecting a productivity topic; \\[4pt]

BT6 & Bot Insisting / Not Satisfied &
Repeating prompts; pushing user toward goal; dismissing or overlooking user resistance. \\[4pt]

BT7 & Context-Aware Responses &
Recognizing user-specific constraints; tailoring suggestions to personal context. \\[4pt]

BT8 & LLM-Led Dialogue Flow &
Driving the conversation; clarifying itself; closing sequences \\[4pt]

BT9 & Opinions on Productivity / Well-Being &
Expressing normative assumptions about productivity or mental health. \\[4pt]

BT10 & Human-Like Social Behaviors &
Using empathic expressions; offering relational cues; performing socially affiliative moves. \\[4pt]

BT11 & Strategies for Self-Reflection &
Asking reflective questions; helping interpret events; guiding users to articulate learnings. \\[4pt]

BT12 & Intrusive or Socially Awkward Behaviors &
Overstepping boundaries; producing emotionally odd, inappropriate, responses. \\

\bottomrule
\end{tabular}
\end{table*}

%TC:endignore
\end{document}